\documentclass{aa}  
\usepackage{graphicx}
\usepackage{txfonts}
\usepackage{multirow}
\usepackage{color}
\usepackage{tabularx}

\usepackage{ulem}

\def\Box{\mathord{\dalemb{7.9}{8}\hbox{\hskip1pt}}}
\def\dalemb#1#2{{\vbox{\hrule height.#2pt
  \hbox{\vrule width.#2pt height#1pt \kern#1pt \vrule width.#2pt}
    \hrule height.#2pt}}}

\def\ba{\begin{eqnarray}}
\def\ea{\end{eqnarray}}
\def\be{\begin{equation}}
\def\ee{\end{equation}}

\def\gtorder{\mathrel{\raise.3ex\hbox{$>$}\mkern-14mu
             \lower0.6ex\hbox{$\sim$}}}
\def\ltorder{\mathrel{\raise.3ex\hbox{$<$}\mkern-14mu
             \lower0.6ex\hbox{$\sim$}}}

\begin{document} 

   \title{Angular distribution of cosmological parameters as a probe of space-time inhomogeneities}
   \author{C. Sofia Carvalho
          \inst{1,2}\fnmsep\thanks{Corresponding author: cscarvalho@oal.ul.pt}
          \and
          Katrine Marques\inst{1}
          }
   \institute{Institute of Astrophysics and Space Sciences, University of Lisbon,
Tapada da Ajuda, 1349--018 Lisbon, Portugal,
         \and
             Research Center for Astronomy and Applied Mathematics, Academy of Athens, Soranou Efessiou 4, 11--527, Athens, Greece 
             }

\abstract{We develop a method based on the angular distribution on the sky of cosmological parameters to probe the inhomogeneity of  large-scale structure and cosmic acceleration. 
We demonstrate this method on the largest type Ia supernova (SN) data set available to date, as compiled by the Joint Light-curve Analysis (JLA) collaboration and, hence, consider the cosmological parameters that affect the luminosity distance. 
We divide the supernova sample into equal surface area pixels and estimate the cosmological parameters that minimize the chi-square of the fit to the distance modulus in each pixel, hence producing maps of the cosmological parameters $\{\Omega_{M}, \Omega_{\Lambda}, H_{0}\}.$ 
In poorly sampled pixels, the measured fluctuations are mostly due to an inhomogeneous coverage of the sky by the SN surveys; in contrast, in well-sampled pixels, the measurements are robust enough to suggest a real fluctuation.
We also measure the anisotropy of the parameters by computing the power spectrum of the corresponding maps of the parameters up to $\ell=3.$  
For an analytical toy model of an inhomogeneous ensemble of homogeneous pixels, we derive the backreaction term in the deceleration parameter due to the fluctuations of $H_{0}$ across the sky and measure it to be of order $10^{-3}$ times the corresponding average over the pixels in the absence of backreaction.  We conclude that, for the toy model considered, backreaction is not a viable dynamical mechanism to emulate cosmic acceleration.}

\titlerunning{Angular distribution of cosmological parameters}
\authorrunning{Carvalho \& Marques}
\maketitle



\section{Introduction}

The best current data, ranging from cosmological to astrophysical scales, have established that the Universe is on average highly homogeneous and isotropic and that it is expanding in an accelerating fashion.  

The observation that, in comparison with nearby sources, distance sources ($z > 0.3$) appear dimmer than predictable in a universe with matter only was interpreted as the need for a dark energy component in the luminosity distance \citep{riess_1998, perlmutter_1999}.
This dimming, however, could be caused by a variation of any of the other components that affect the luminosity distance, namely by inhomogeneities in the energy densities or anisotropies in the expansion \citep{amendola_2013}.
On large scales, the matter distribution is statistically homogeneous. Locally, however, matter is distributed according to a pattern of alternate overdensed regions (filaments) and underdensed regions (voids) \citep{scrimgeour_2012,watson_2013}, which is thought to reflect quantum fluctuations away from homogeneity that originated in the very early Universe. 
In this paper, we hypothesize that the resulting inhomogeneity in the matter distribution might be imprinted in an angular distribution of the cosmological parameters, which is measured at the scale of the matter distribution pattern, about the average parameter values. 

Inhomogeneities in the cosmological parameters distribution might create a dynamical mechanism that would emulate a cosmic acceleration as a derived, instead of fundamental, effect (see e.g. \citet{matarrese_2006,larena_2008,clarkson_2011,chiesa_2014,adamek_2014}). 
Conversely, inhomogeneities in the cosmological parameters might reflect new physics, such as a dynamical dark energy component or a modification to the Einstein's theory of general relativity (see e.g. \citet{amendola_2013}).  On the one hand, inhomogeneities in the matter distribution might generate fluctuations in the dynamical dark energy, which would imply angular variations of the Hubble parameter across the sky that could be observed as anisotropies of the luminosity distance as a function of redshift $d_{L}(z);$ on the other hand, fluctuations in the new gravitational fields could have imprints on the local expansion and, consequently, on $d_L(z)$ \citep{cooray_2010}.

Supernovae establish the relation of the luminosity distance with redshift up to $z\approx 1.$ The luminosity distance is sensitive to both the geometry and growth of structure \citep{bonvin_2006},  
which renders supernovae 
a sensitive probe of the late-time expansion history of the Universe.
The current amount of data on SNe from the combination of different SN surveys, if large and homogeneous enough, could be used to investigate the angular distribution of inhomogeneities and hence to investigate the nature of cosmic acceleration. 

Supernovae have previously been used to constrain inhomogeneities in the cosmic expansion, namely by measuring the hemispherical anisotropy of $H_{0}$ \citep{kalus_2013}, assessing the dependence of $H_{0}$ with the position on the cosmic web \citep{wojtak_2014}, mapping $\{H_{0}, q_{0}\}$ in the sky \citep{bengaly_2015}\footnote{We learned about this article while the work reported here was in progress.}, and by estimating the statistical significance of rejecting an isotropic magnitude--redshift relation \citep{javanmardi_2015}\footnote{We learned about this article while the work reported here was under review.}. Supernovae have also been used to constrain inhomogeneities in the dark energy, namely by constraining fluctuations of dark energy \citep{blomqvist_2010,cooray_2010} and by constraining radial inhomogeneity for a Lema\^itre--Tolman--Bondi metric with a cosmological constant \citep{marra_2014}.

In this paper, we develop a method to probe inhomogeneities in the form of a local estimation of cosmological parameters. We demonstrate the method on type Ia supernovae, in particular by dividing the supernova sample into different regions of the sky and estimating the angular distribution of the cosmological parameters that affect the luminosity distance, namely $\{\Omega_{M}, \Omega_{\Lambda}, H_{0}\},$ across the different regions.Each region is assumed to be described by a Friedmann--Lema\^itre--Roberston--Walker metric so that the full sky is an inhomogeneous ensemble of disjoint, locally homogeneous regions. This ensemble of homogeneous regions is an approximate solution to
the Einstein equation, which we assume to be a sufficient approximation to enable us to interpret the cosmological parameters $\{\Omega_{M}, \Omega_{\Lambda}, H_{0}\}$ as a description of the true energy content. 
We also suggest measuring the anisotropy in the angular distribution of the parameters not by looking for particular asymmetries at particular scales, but by computing the angular power spectrum that includes all multipoles accessible in the regions of the sky. We then use the angular fluctuations in $H_{0}$ to estimate the backreaction in the sense of Buchert.

Since the Einstein tensor is non-linear in the metric, the average evolution of an inhomogeneous space-time is not the same as the evolution of a smooth space-time with the same initial conditions \citep{buchert_1999}. The effect of inhomogeneities on the average properties is known as backreaction. 
When averaging over the homogeneous regions, angular fluctuations in the expansion factor, and consequently in $H_{0}$, induce a backreaction term in the form of an extra positive acceleration \citep{rasanen_2006}. We derive the analytical extra positive acceleration for a toy model of an arbitrary number of disjoint, homogeneous regions and compute the overall deceleration parameter assuming a) no backreaction and b) backreaction for the measured angular distribution of $H_{0}.$ 

This paper is organized as follows. In Sec.~\ref{sec:sn} we describe the SN data and the theory for the calculation of the observables.  In Sec.~\ref{sec:param_est} we introduce the method by performing both a global and a local estimation, according to the parameter estimation described in App.~\ref{sec:est_method}. 
From the global estimation, there result parameter values taken for fiducial values. 
From the local estimation, there result maps of the cosmological parameters.  
In Sec.~\ref{sec:test} we include three tests of the inhomogeneity of the SN surveys. The first test is in the SN distribution in redshift, the second test is in the SN distribution in both redshift and sampling, and the third test is in the removal of a noise bias hypothesized as a measure of the inhomogeneity of the SN sampling. We then proceed to analyse the results. In Sec.~\ref{sec:power_spec} we compute the power spectrum of the maps of the parameters. In Sec.~\ref{sec:backreact}, for a toy model of backreaction, we compute the values of the parameters averaged over the pixels. 
These are, to the best of our knowledge, the first estimates of inhomogeneities from the angular distribution of cosmological parameters and the first estimates of backreaction effects. In Sec.~\ref{sec:concl} we discuss current limitations of the method owing to insufficient data and we indicate future research directions.

\begin{figure}
\centerline{
\includegraphics[width=\columnwidth]
{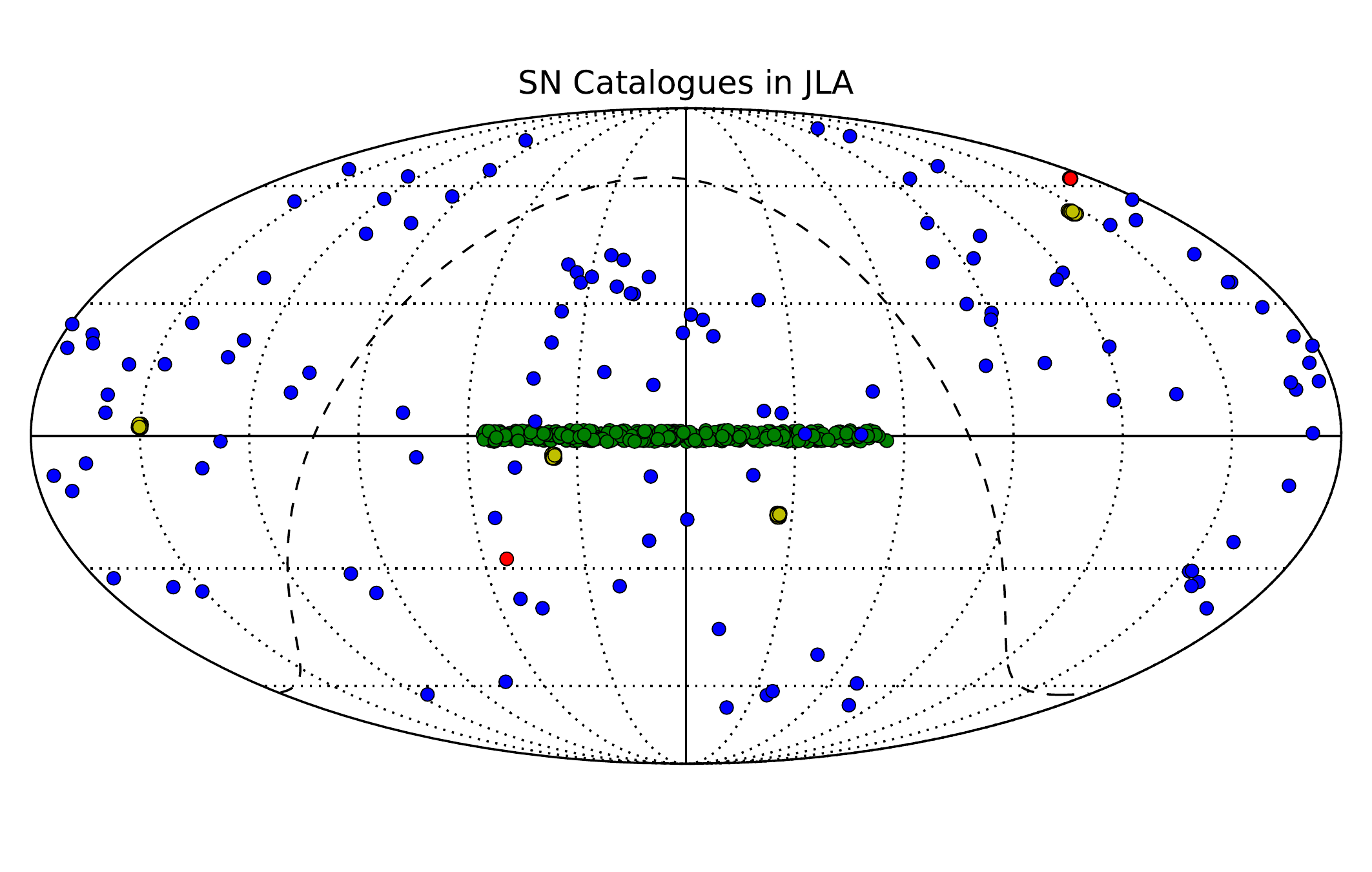} 
}
\vspace{-0.3cm}
\caption{
{\bf Angular distribution of the JLA type Ia SN sample in celestial coordinates.} The sample consists of type Ia SNe compiled from different surveys as described in \citet{betoule_2014}. The blue dots indicate  low--z data. The green dots indicate SDSS--II data. The yellow dots indicate SNLS data. The red dots indicate HST data. The dashed line represents the Galactic equator.
}
\label{fig:jla_sn}
\end{figure}

\begin{figure*}
\centerline{
\includegraphics[width=10cm]
{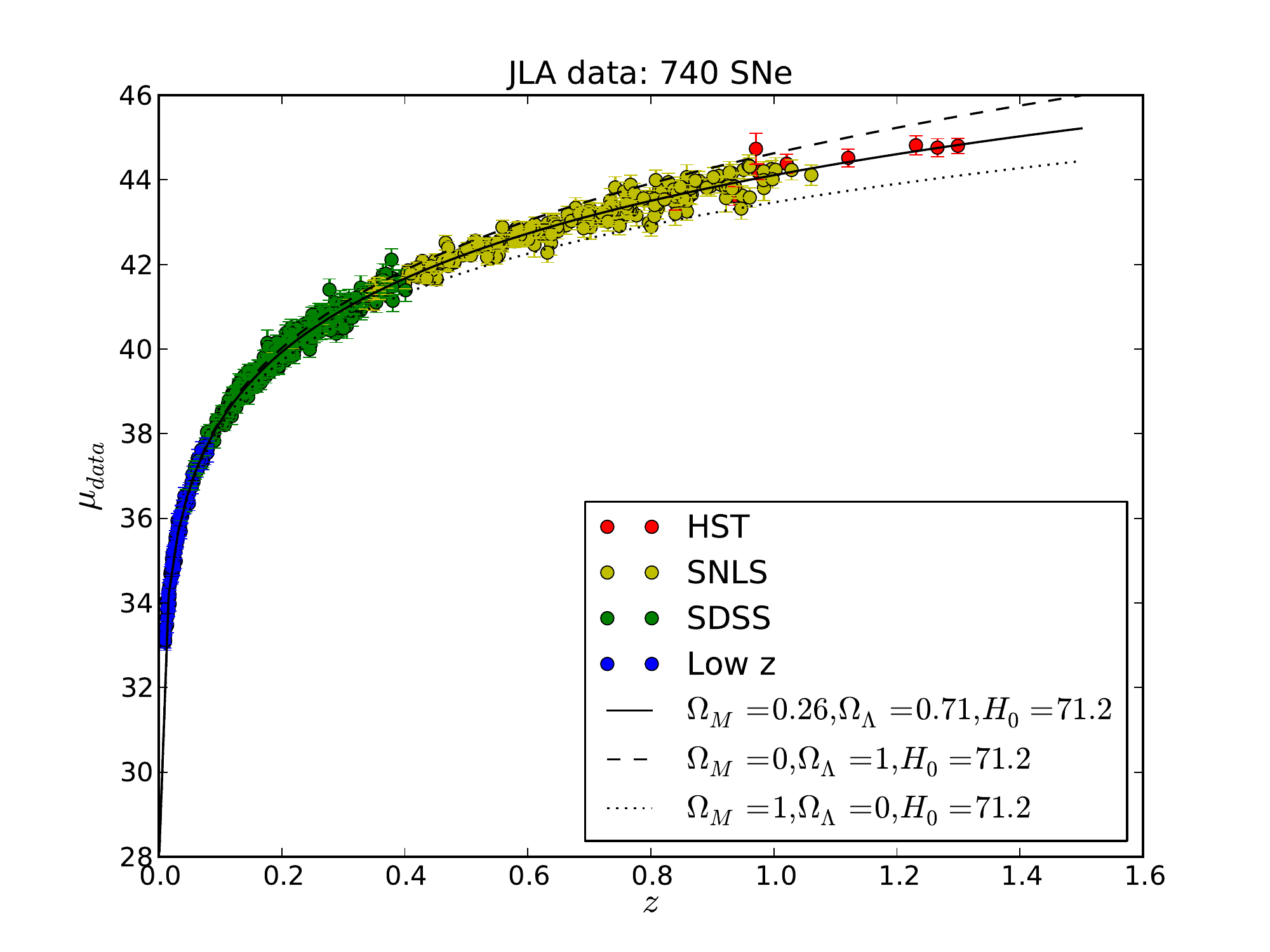}
\includegraphics[width=10cm]
{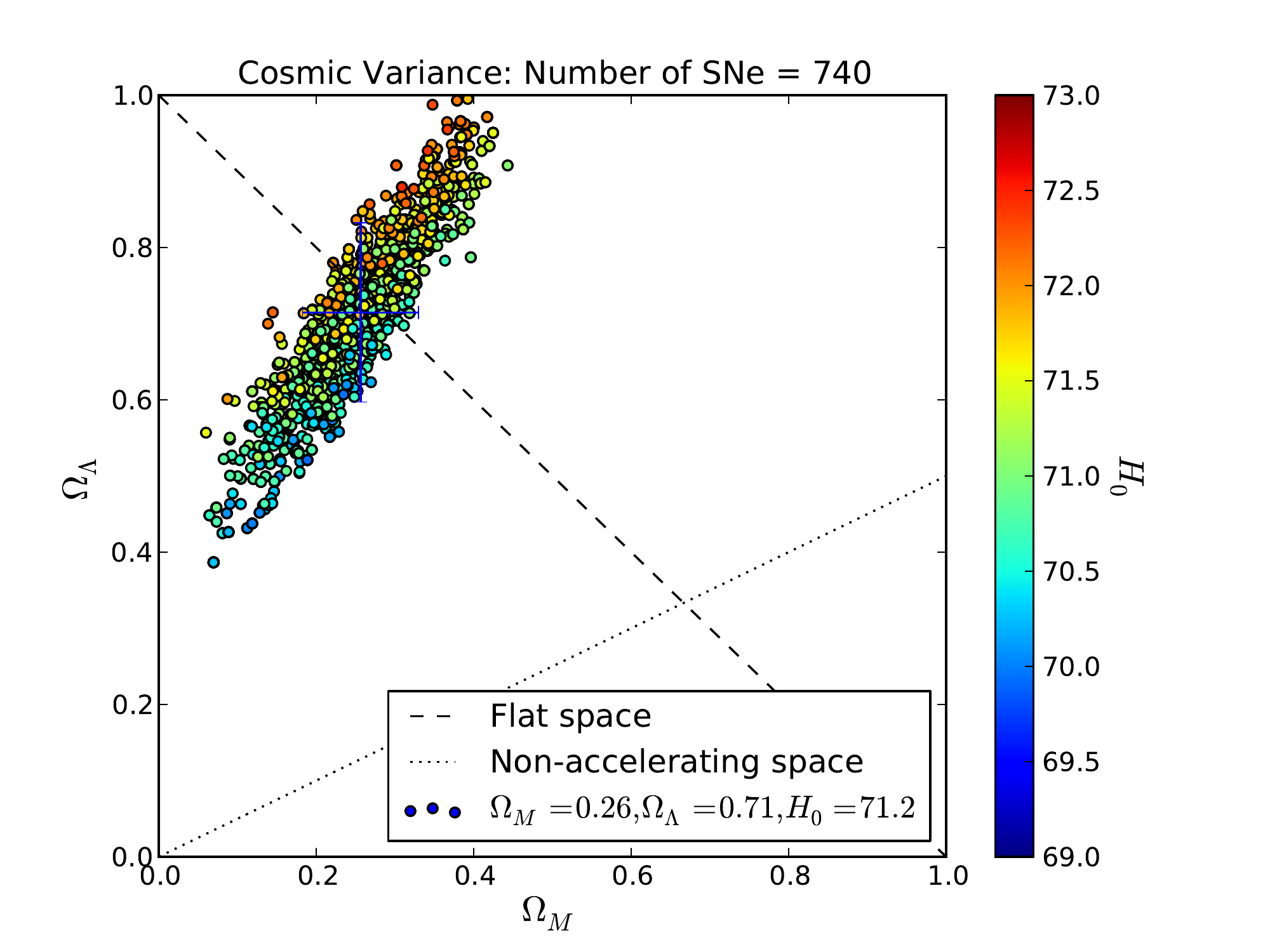}
}
\vspace{-0.3cm}
\caption{
{\bf Left panel: distance modulus as a function of redshift.} 
The dots indicated in colour according to the SN survey and indicate the distance modulus $\mu_{\text{data}}$ computed from the data on each SN in the sample. The black line indicates the distance modulus $\mu_{\text{theo}}$ computed theoretically for the values of $\{\Omega_{M},\Omega_{\Lambda}, H_{0}\}$  estimated from the complete sample. The black dashed and black dotted lines indicate the distance modulus for two extreme cases: $\Omega_{M}=0,\Omega_{\Lambda}=1$ and $\Omega_{M}=1,\Omega_{\Lambda}=0,$ respectively.
{\bf Right panel: Markov chain of one realization.}
The dots indicate the entries in the Markov chain. The position in the graph indicates the values of $(\Omega_{M},\Omega_{\Lambda})$ of each entry according to the axes, whereas the coloured dots indicate the value of $H_{0}$ according to the side colour bar. The dashed line represents $\Omega_{M}+\Omega_{\Lambda}=1$ and the dotted line represents $\Omega_{M}/2-\Omega_{\Lambda}=0.$
}
\label{fig:jla_mu_z}
\end{figure*}

\section{Supernova data and theory}
\label{sec:sn}

We use the type Ia supernova sample compiled by the Joint Light--curve Analysis (JLA) SNLS--SDSS collaborative effort \citep{betoule_2014}. 
The JLA collaboration  aimed to include SNe from surveys at different redshifts, improve the accuracy of the calibration, and make the light--curve model uniform across the surveys. 
The combined sample consists of type Ia SNe from different supernova surveys, namely
the Sloan Digital Sky Survey II (SDSS--II) at intermediate redshifts \citep{sako_2014} and the Supernova Legacy Survey (SNLS) combined with Hubble Space Telescope (HST) surveys at high redshifts and other surveys at low redshifts \citep{conley_2011}, totalling $N_{\text{SNe}}=740$ SNe with redshift $0.01 \le z \le 1.30.$ The sample includes the SNe in the Hubble flow ($z \ge 0.01$) that are not affected by strong Milky Way contamination. The sample was obtained from {\tt http://supernovae.in2p3.fr/sdss\_snls\_jla/ReadMe.html}. The SNe are distributed on the sky according to Fig.~\ref{fig:jla_sn}.  

From the sampled data, we compute the distance modulus for each SN as
\ba
\mu_{\rm data}(z)=m_{B}(z)-M_{B,\rm eff}.
\label{eqn:mu_data}
\ea
For each SN we take the peak magnitude $m_{B}$ inferred from the observed magnitude and corrected for the selection bias, and we define the effective absolute magnitude $M_{B,\rm eff}$ as the absolute magnitude $M_{B}$ corrected by the time stretching of the light curve $x_{1}$ and the supernova colour at maximum brightness $c$ as follows:
\ba
M_{B,\rm eff}=M_{B}-\alpha x_{1}+\beta c.
\label{eqn:M_B_eff}
\ea
The parameters $M_{B}=-19.05\pm 0.02,$ $\alpha=0.141\pm 0.006,$ and $\beta=3.101\pm 0.075$ are estimated for all SNe, whereas the parameters $\{m_{B},x_{1}, c\}$  are estimated for each SN separately by fitting a model of the Ia SN spectral sequence to the photometric data. 
We compute $\mu_{\rm data}$ for all SNe in the sample and plot it in the left panel of Fig.~\ref{fig:jla_mu_z}. The error 
$\sigma_{\mu_{\text{data}}}$ results from the error propagation according to Eqs.~(\ref{eqn:mu_data}) and (\ref{eqn:M_B_eff}) assuming uncorrelated errors. 

We now establish the dependence of the distance modulus with the cosmological parameters.
The distance modulus is defined in terms of the luminosity distance $d_{L}$ (measured in Mpc) as
\ba
\mu_{\rm theo}(z;\Omega_{M},\Omega_{\Lambda},H_{0})
=5\log[d_{L}(z;\Omega_{M},\Omega_{\Lambda},H_{0})]+25.
\ea
For a (3+1)-dimensional line  element expressed in spherical coordinates as
\ba
ds^2=-c^2 dt^2+a^2(t)\left[ {dr^2\over {1-\kappa r^2}}+r^2 d^2\Omega_{(2)}\right],
\ea
where $\kappa=\{1,0,-1\}$ characterizes the spatial curvature and $d^2\Omega_{(2)}\equiv d^2\theta+\sin^2(\theta)d^2\phi$ stands for the two-sphere line element, the luminosity distance is given by
\ba
d_{L}(z)={a_{0}(1+z)\over \sqrt{|\kappa|}}S_{\kappa}\left[
  {c\sqrt{|\kappa|}\over{a_{0}H_{0}}}\int_{0}^{z}{dz\over E(z)}
\right],
\ea
where 
\ba
S_{\kappa}(x)=
\begin{cases} 
\sin(x), & \text{if $\kappa >0$}\\
x, & \text{if $\kappa =0$}\\
\sinh(x), & \text{if $\kappa <0$}
\end{cases}
\ea
and 
\ba
E(z)
=\left[ \Omega_{M}(1+z)^3+\Omega_{\kappa}(1+z)^2+\Omega_{\Lambda}\right]^{1/2}
,\ea
with the dimensionless energy density of matter, curvature, and dark energy at present time given, respectively, by
\ba 
\Omega_{M}={8\pi G_{N} \over 3H_{0}^2}\rho,\quad 
\Omega_{\kappa}=-{{c^2\kappa}\over a_0^2H_0^2},\quad
\Omega_{\Lambda}={\Lambda\over 3H_{0}^2}.
\ea
Assuming that
\ba
\Omega_{M}+\Omega_{\kappa}+\Omega_{\Lambda}=1,
\label{eqn:flatness}
\ea
then
\ba
&&d_{L}(z)
=
{c(1+z)\over \sqrt{|\Omega_{\kappa}|}H_0} S_{k}\Biggl[
    \sqrt{|\Omega_{\kappa}|}\times\cr
&\times&\int_{0}^{z} {dz^{\prime}\over \left[ 
      (1+\Omega_{M}z^{\prime})(1+z^{\prime})^2-\Omega_{\Lambda}(2+z^{\prime})z^{\prime}
    \right]^{1/2}}
  \Biggr].
\label{eqn:dl}
\ea
Hence the distance modulus via the luminosity distance depends on the following set of cosmological parameters $\{\Omega_{M},\Omega_{\Lambda}, H_{0}\} $ subject to the condition in Eq.~(\ref{eqn:flatness}).

\begin{table*}
\caption{
{\bf Values for the parameters estimated from the JLA type Ia SN sample.} 
}
\label{table:param}
\centering
\begin{tabular}{c|cc|ccc}
\hline\hline
Parameter 
& \multicolumn{2}{c|} {Complete sample} 
& JLA Comb & WMAP 9yr & Planck \\ 
& All & Flat & \citep{betoule_2014}  & \citep{hinshaw_2013} & \citep{planck_2013} \\ \hline
$\Omega_{M}$ 
& $0.256\pm 0.074$ & $0.267\pm 0.018$ 
& $0.305\pm 0.010$ & $0.233\pm 0.023$ & $0.263\pm 0.007$ \\
$\Omega_{\Lambda}$ 
& $0.715\pm 0.118$ & $0.733\pm 0.018$ 
& $0.697\pm 0.010$ & $0.721\pm 0.025$ & $0.689\pm 0.020$ \\
$H_{0}$ 
& $71.17\pm 0.44$ & $71.21\pm 0.33$ 
& $68.3\pm 1.03$ & $70.0\pm 2.2$ & $67.4\pm 1.4$ \\
\hline
$\Omega_{\kappa}$ 
& $0.029\pm 0.140$ & $0.000\pm 0.026$
& $0.002\pm 0.003$ & $0.046\pm 0.034$ & $0.051\pm 0.007$ \\
$q_0$ 
& $-0.586\pm 0.124$ & $-0.599\pm 0.020$ 
&  $-0.541\pm 0.012$ & $-0.604\pm 0.028$ & $-0.554\pm 0.004$ \\
\hline
\end{tabular}
\tablefoot{
Column 1: the parameters estimated either directly or indirectly from the data. Columns 2--3: the values estimated from the complete SN sample, both when allowing $\{\Omega_{M}, \Omega_{\Lambda}, H_{0}\}$ to vary and when imposing $\Omega_{M}+\Omega_{\Lambda}=1.$ 
Columns 4--6: the values estimated by other collaborations from SN data, CMB data, and combined data.
}
\end{table*}

\section{Method}
\label{sec:param_est}

We introduce the method by performing the estimation of the parameters $\{\Omega_{M},\Omega_{\Lambda}, H_{0}\}$ as described in Appendix~\ref{sec:est_method}, first from the complete SN sample and then from SN subsamples grouped by regions in the sky.

\subsection{Global parameter estimation}

We first use the complete SN sample to fit $\mu_{\text{theo}}$ to $\mu_{\text{data}}$ and estimate the maximum-likelihood values for the parameters $x_i=\{\Omega_{M},\Omega_{\Lambda}, H_{0}\}.$ Whenever unspecified, $H_{0}$ is measured in units of $\text{km}~\text{s}^{-1}\text{Mpc}^{-1}.$  
We ran ten realizations of Markov chains of length $10^{4}.$ The starting point of each realization is randomly generated. 
We remove the first $20\%$ of entries in the chains to keep the burnt-in phase only. We then thin down the remaining $80\%$ of entries to a half by removing one of each consecutive entry to eliminate correlations within each chain. Averaging over the various realizations of Markov chains, we obtain the following results: 
$x^{\text{fid}}_{i}\equiv \{\Omega_{M}^{\text{fid}},\Omega_{\Lambda}^{\text{fid}}, H_{0}^{\text{fid}}\}=
\{0.256,0.715,71.17\}\pm\{0.074,0.118,0.44\}.$
If we moreover impose 
$\Omega_{M}+\Omega_{\Lambda}=1,$ thus enforcing  
$\Omega_{\kappa}=0,$ then we obtain the following results: 
$x^{\text{fid}}_{i,\text{flat}}\equiv \{\Omega_{M}^{\text{fid}},\Omega_{\Lambda}^{\text{fid}}, H_{0}^{\text{fid}}\}_{\text{flat}}=
\{0.267,0.733,71.21\}\pm\{0.018,0.018,0.33\}.$ (See Table \ref{table:param}.) 
These are the mean values $x_{i}^{\text{fid}}$ and $x_{i,\text{flat}}^{\text{fid}},$ that we use as fiducial values in the subsequent calculations. The errors contain the dispersion in each chain and the dispersion among the averages of the different chains added in quadrature. 
By fixing $H_{0}=70.0~\text{km~s}^{-1}\text{Mpc}^{-1}$ and imposing $\Omega_{M}+\Omega_{\Lambda}=1,$ we obtain $\Omega_{M, \text{flat,} H_{0}=70.0}=0.322\pm 0.012.$ 

We recall that the JLA collaboration found $\Omega_{M,\text{JLA}}=0.295\pm0.034$ for a fixed Hubble parameter  $H_0=70.0~\text{km~s}^{-1}\text{Mpc}^{-1},$ which includes our result $\Omega_{M, \text{flat,}H_{0}=70.0}$ within the errors. 
The results obtained by the JLA collaboration in combination with other data were 
$\{\Omega_{M},\Omega_{\Lambda},H_{0}\}_{\text{JLAComb}}=\{0.305,0.697,68.3\}\pm \{0.010,0.010,1.03\}$ \citep{betoule_2014}.
For comparison, in Table \ref{table:param} we also include the values for these parameters obtained from the ninth-year WMAP data \citep{hinshaw_2013} and the Planck collaboration \citep{planck_2013}.
We observe that our results are compatible with the WMAP results, but favour lower values for $\Omega_{M},$ and simultaneously higher values for $\Omega_{\Lambda}$ and for $H_{0},$ in comparison to the JLAComb and Planck results. 
In contrast, Riess et al. found $H_0=73.8\pm2.4~\text{km~s}^{-1}\text{Mpc}^{-1}$ \citep{riess_2011}.

A derived parameter of interest is the curvature energy density $\Omega_{\kappa},$ which  becomes a dependent variable owing to the flatness condition in Eqn.~(\ref{eqn:flatness}). The fiducial values return $\Omega_{\kappa}^{\text{fid}}=0.029\pm 0.140$ and $\Omega_{\kappa,\text{flat}}^{\text{fid}}=0.000\pm0.026.$ The values for $\Omega_{\kappa}$ returned by the above-mentioned studies are included in Table \ref{table:param}.
We observe that our results are compatible with the WMAP results but favour values for $\Omega_{\kappa}$ that are higher than the
JLA Comb results and lower than the Planck results.

Another derived parameter of interest is the cosmic deceleration parameter $q_{0},$ which is defined in terms of $\{\Omega_{M}, \Omega_{\Lambda}\}$ as 
\ba
q_{0}={\Omega_{M}\over 2}-\Omega_{\Lambda}=-{1\over H_{0}^2}{\ddot a \over a}.
\ea
The fiducial values return $q_{0}^{\text{fid}} =-0.586\pm 0.124$ and $q_{0,\text{flat}}^{\text{fid}} =-0.599\pm 0.020.$ The values for $q_0$ returned by the above-mentioned studies are included in Table \ref{table:param}.
We observe that our results are compatible with the WMAP result but favour lower values for $q_{0}$ in comparison to the JLAComb and Planck results.

We compute $\mu_{\rm theo}$ for the estimated values $x_{i}^{\text{fid}}$ and plot it as a black line in the left panel of Fig.~\ref{fig:jla_mu_z}. 
In the right panel of Fig.~\ref{fig:jla_mu_z}, we plot the Markov chain of one realization. We observe that a) the points are aligned approximately parallel to the lines of constant $H_{0} $ and b) the points form approximately an ellipse whose largest axis is not parallel to the lines of constant $q_{0}.$

To assess the convergence, we apply the Gelman-Rubin heuristic convergence test. For each parameter $x_i$ and  $N_{\text{chain}}$ chains, we compute the mean $x_{ij}$ and variance $\sigma_{x_{ij}}^2$ of each chain $j.$ We take the mean of all chains $\bar x_{i}=\big<x_{ij}\big>_{j}$ and compute the variance of the means of the chains  $\sigma^2_{\bar x_{i}}=\sigma^2(x_{ij}~\text{about}~\bar x_{i}).$ We compute the average of the variances of the chains  $\bar \sigma^2_{i}=\big<\sigma^2_{x_{ij}}\big>_{j},$ and compare with $\sigma^2_{\bar x_{i}}$ forming the ratio
\ba
r_{GR}={
{{N_{\text{chain}}-1}\over N_{\text{chain}}}\sigma^2_{\bar x_{i}}+{1\over N_{\text{chain}}}\bar \sigma^2_{i}
\over 
\sigma^2_{\bar x_{i}} }.
\ea
If the chains are well mixed and have all sampled the parameter space well, then $r_{GR}\sim 1.$ We find $r_{GR}\approx 0.90$ for the estimation of both $x_{i}^{\text{fid}}$ and $x_{i,\text{flat}}^{\text{fid}}.$

\begin{figure}[t]
\centerline{
\includegraphics[width=\columnwidth]
{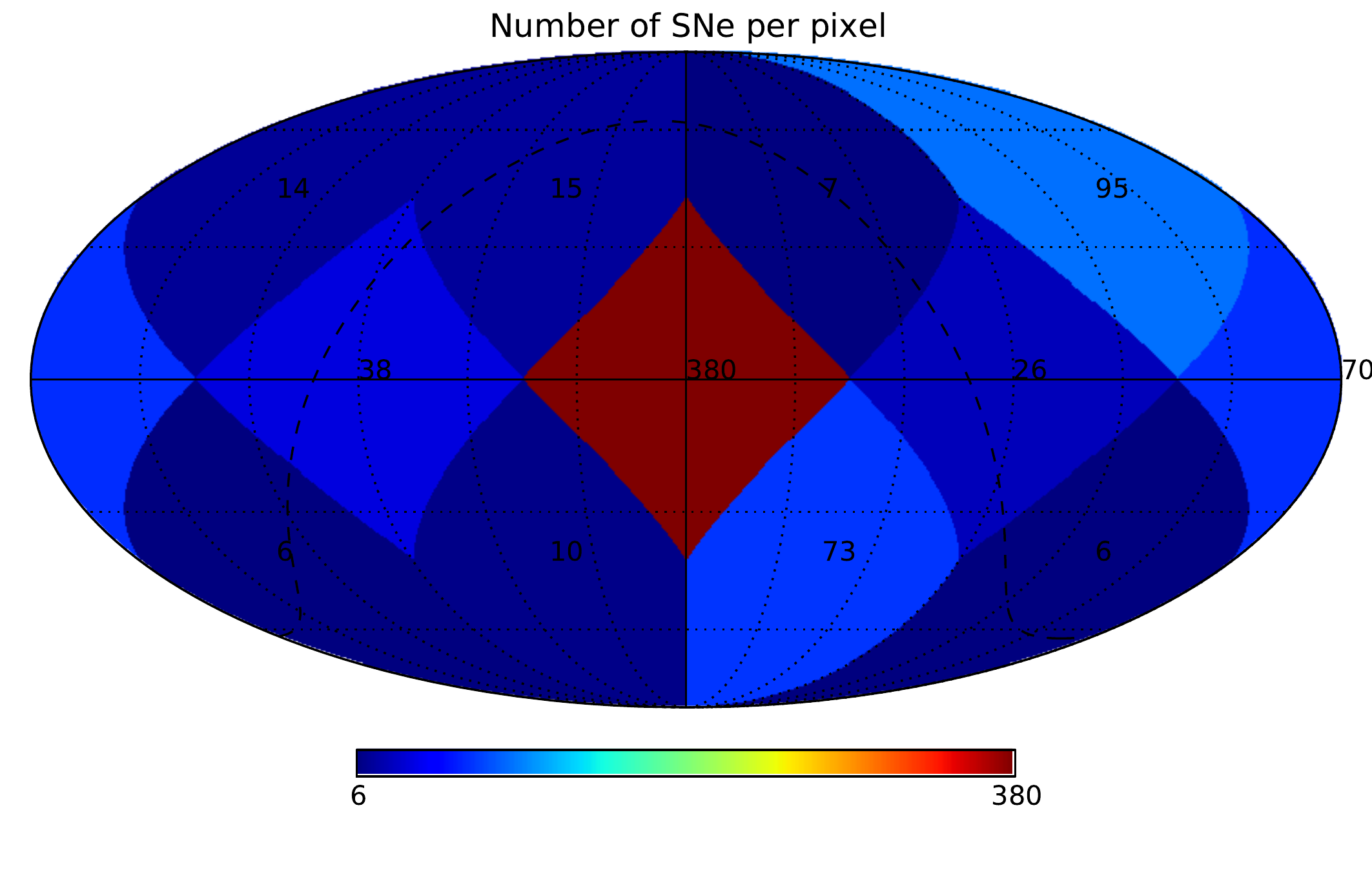}
}
\vspace{-0.3cm}
\caption{
{\bf Angular distribution of the number of type Ia SNe per pixel.} The total SN sample is divided into subsamples grouped by pixels in the HEALPix pixelation scheme. The number of SNe in each pixel is printed in the centre of the pixel. The pixel indices $\{1,2,3,4,5,6,7,8,9,10,11,12\}$ correspond to the SN counts $\{15, 14, 95, 7, 380, 38, 70, 26, 10, 6, 6, 73\}.$ The dashed line represents the Galactic equator.
}
\label{fig:sn_per_pix}
\end{figure}

\begin{figure*}
\centerline{
\includegraphics[width=9cm]
{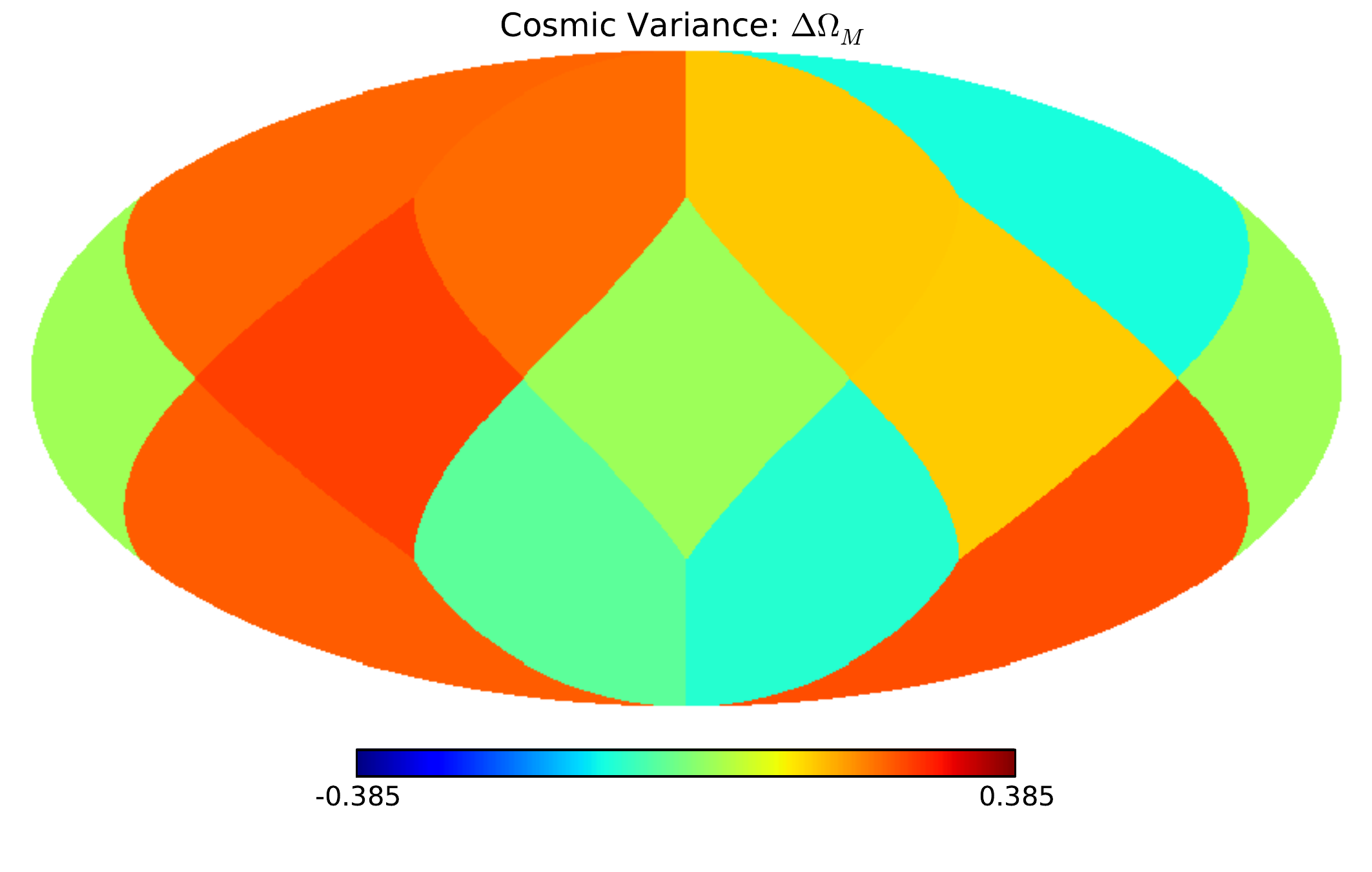}
\includegraphics[width=9cm]
{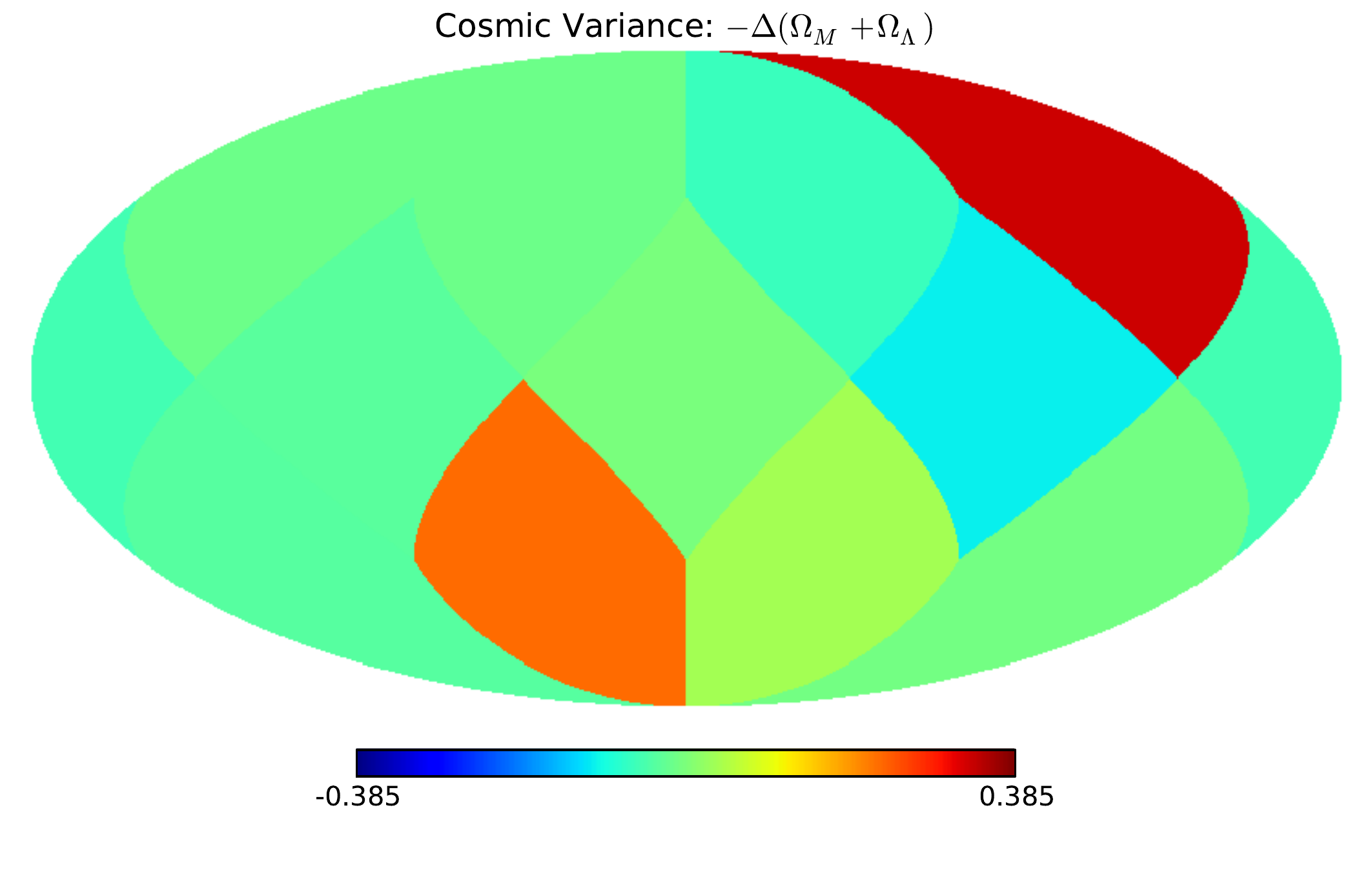}
}
\centerline{
\includegraphics[width=9cm]
{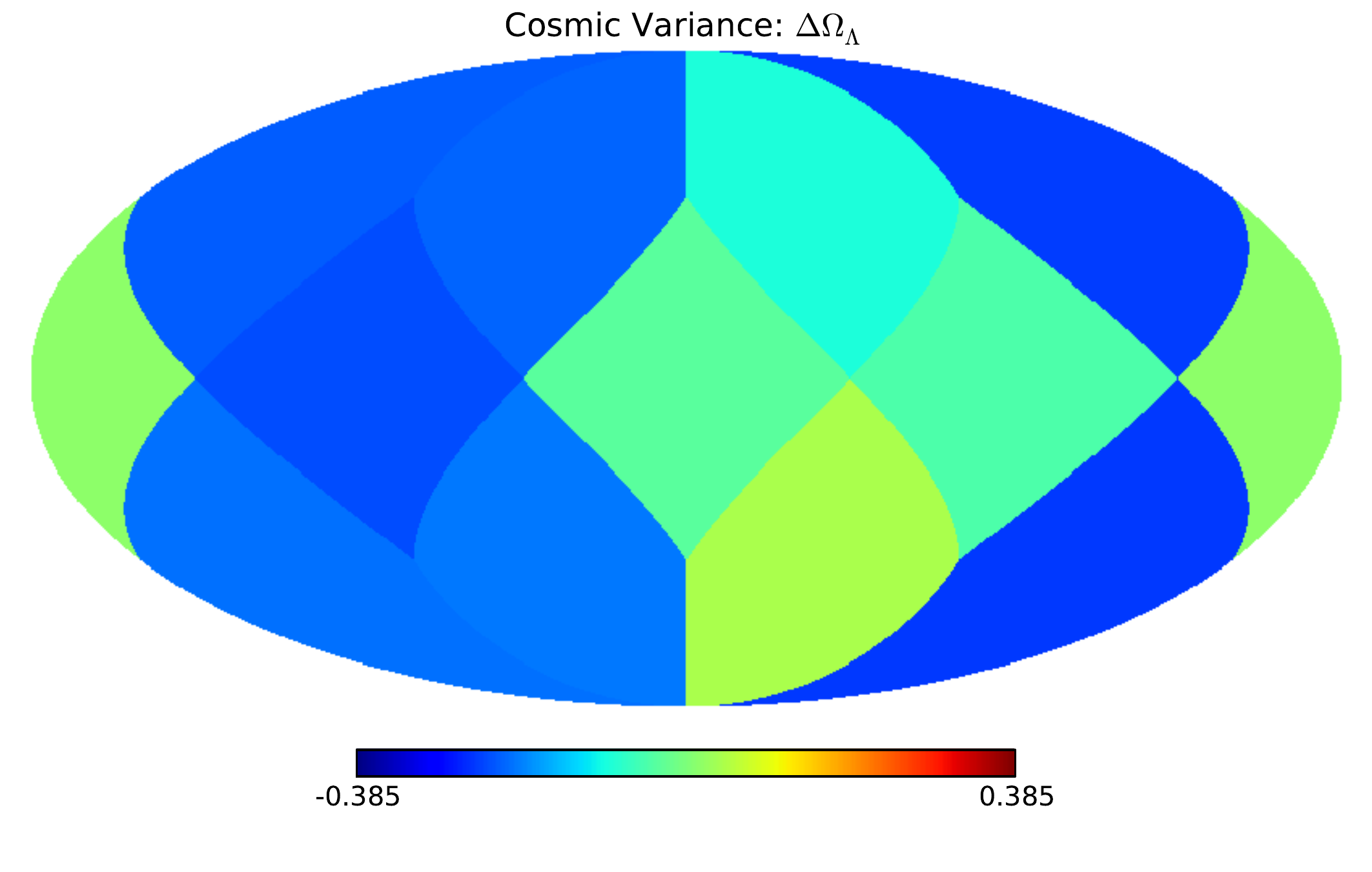}
\includegraphics[width=9cm]
{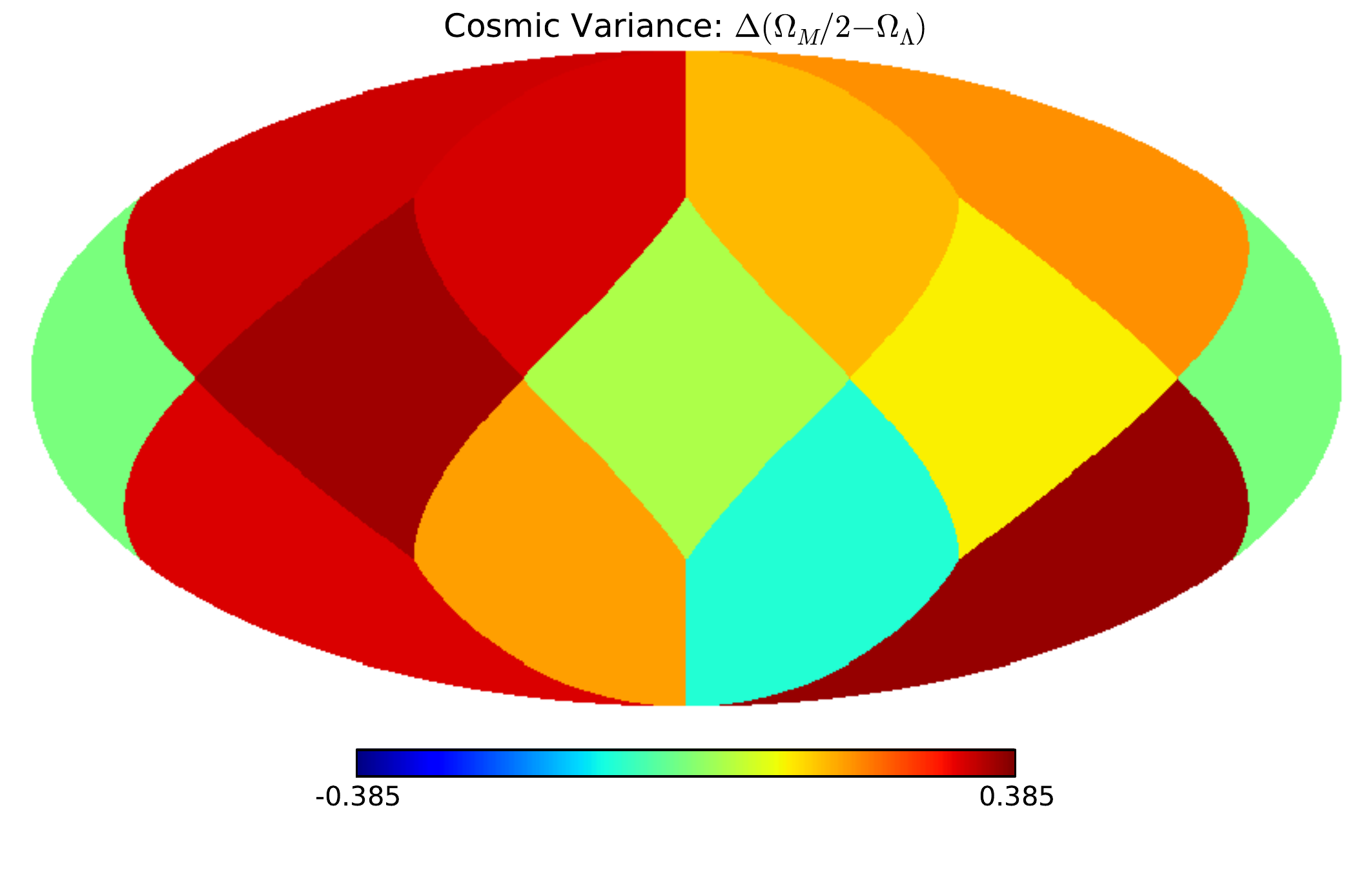}
}
\centerline{
\includegraphics[width=9cm]
{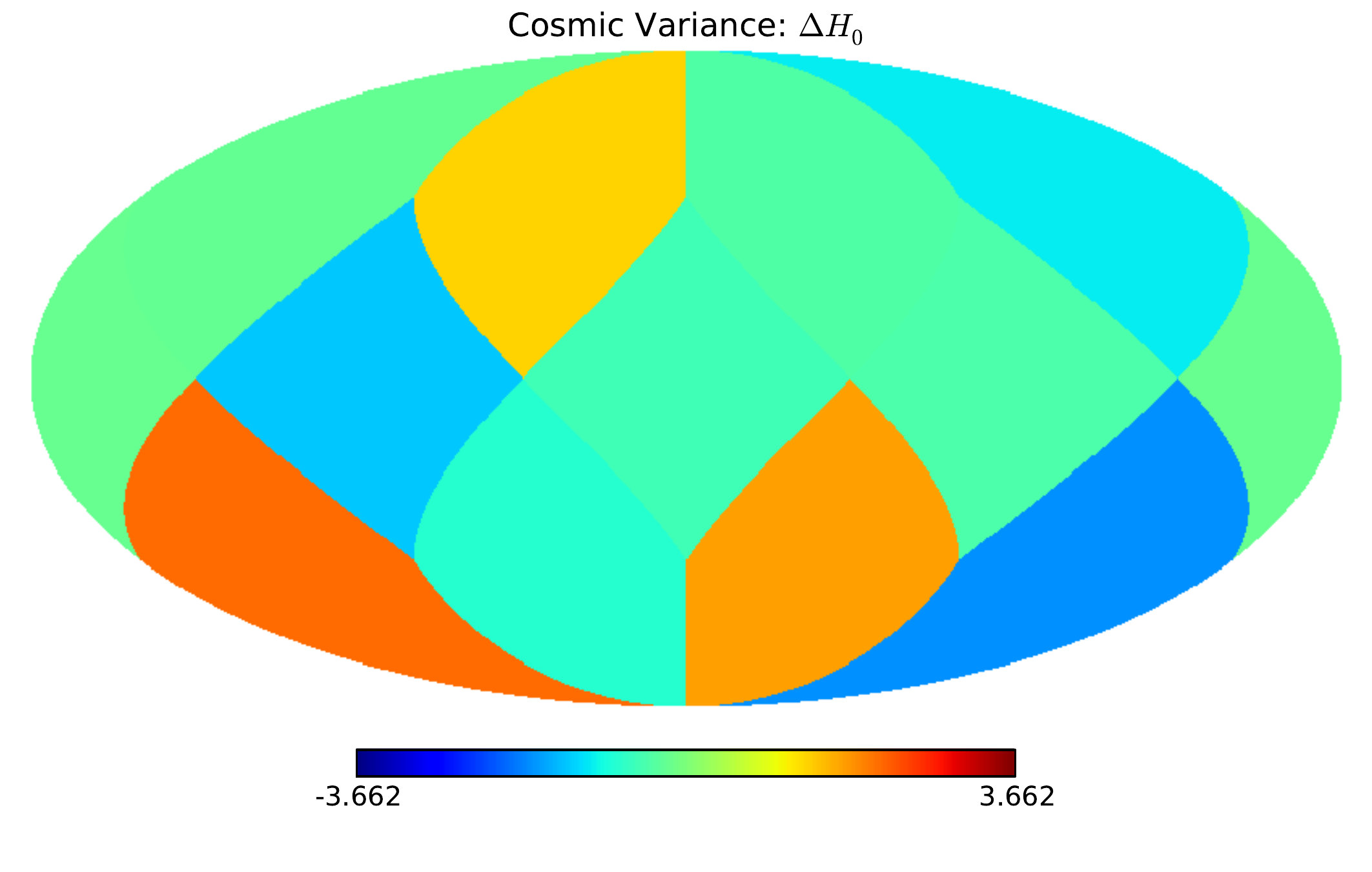}
}
\vspace{-.3cm}
\caption{
{\bf Angular distribution of the estimated parameters $\{\Omega_{M},\Omega_{\Lambda}, H_{0}\}_{k,\text{cosmic\_var}}$ and the derived parameters $\Omega_{\kappa}$ and $q_{0}$ at each pixel $k.$} The subsample of type Ia SNe in each pixel is used to estimate the value of the parameters $\{\Omega_{M},\Omega_{\Lambda}, H_{0}\}$ in that pixel, resulting in a map for each parameter.  Operating the estimated maps, results in the derived maps for $\Omega_{\kappa}$ and $q_{0}.$  Subtracting  the fiducial values, results in corresponding difference maps. 
The fiducial values are $\{\Omega_{M}^{\text{fid}},\Omega_{\Lambda}^{\text{fid}}, H_{0}^{\text{fid}}\}=\{0.256,   0.715,   71.17\},$ $\Omega_{\kappa}^{\text{fid}}=0.029,$ and $q_{0}^{\text{fid}}=-0.586.$
}
\label{fig:map_param_cosmic_var}
\end{figure*}

\subsection{Local parameter estimation: Cosmic variance}

We proceed to divide the SNe into subsamples grouped by regions in the sky. From each subsample, we aim to estimate the parameters $x_i=\{\Omega_{M},\Omega_{\Lambda}, H_{0}\}$ and produce maps such that the values in each region are given by the values of the parameters estimated with the subsample in that region. Hence we must guarantee that the regions have the same surface area and all contain SNe. Ideally each pixel should also have a SN subsample that covers a range of $z$ enough to break the degeneracies among the parameters 
in Eq.~(\ref{eqn:dl}). We use the HEALPix pixelation scheme to divide the sphere into pixels of equal surface area \citep{gorski_2005}, hence defining the regions as pixels. In the HEALPix pixelation scheme, we select ${\tt nside}=1,$ which  produces a map divided into ${\tt npixel}=12$ pixels. The number of SNe in each pixel is indicated in Fig.~\ref{fig:sn_per_pix}. In this description, the complete sample is equivalent to ${\tt npixel}=1$ pixel. 

We then use the SN subsample in each pixel to estimate the maximum--likelihood values for the parameters $\{\Omega_{M},\Omega_{\Lambda}, H_{0}\}.$ This estimation is performed, as in the case of the complete sample, on realizations of the original SN distribution in redshift and in space, i.e. for the original dependence of redshift with the position and for the original sampling per pixel. The variance of the estimation in each pixel determines the accuracy with which we can measure the parameters with this sample. For this reason we call this estimation ``Cosmic variance.'' In each pixel, we ran 30 realizations of Markov chains of length $10^4.$ In each pixel, the starting point of each realization is randomly generated as for the case of the complete sample. Also, as in the complete sample, we perform the burnt-in selection and the thinning reduction of the Markov chains in each pixel. Hence from each chain $j,$ at each pixel $k,$ we obtain an estimate for each parameter $x_{i}$. The parameter estimation results in maps $x_{ijk}=\{\Omega_{M},\Omega_{\Lambda}, H_{0}\}_{jk}.$ We average the maps of each parameter over the chains by taking the mean weighted by the number of SNe in each pixel, $\bar x_{ik}=\big< x_{ijk}\big>_{j}.$ Since the number of SNe in each pixel is the same for the different chains, this averaging is equivalent to the simple mean of the maps by pixel. We denote the resulting mean maps of the parameters by $\bar x_{ik}=\{\Omega_{M},\Omega_{\Lambda}, H_{0}\}_{k,\text{cosmic\_var}}.$ 
To assess the convergence, we compute the Gelman--Rubin ratio, finding $r_{GR}\approx 0.90$ for the estimation of all parameters at all pixels.

For better visualization, we compute difference maps $\Delta \bar x_{ik}=\bar x_{ik}-x_{i}^{\text{fid}}$ by subtracting the fiducial values $\{\Omega_{M}^{\text{fid}},\Omega_{\Lambda}^{\text{fid}}, H_{0}^{\text{fid}}\}$ off the corresponding maps $\{\Omega_{M},\Omega_{\Lambda}, H_{0}\}_{k,\text{cosmic\_var}}.$ 
Averaging over the various realizations of Markov chains at each pixel, we obtain the difference maps $\Delta \bar x_{ik}=\{\Delta\Omega_{M},\Delta\Omega_{\Lambda}, \Delta H_{0}\}_{k,\text{cosmic\_var}}$ shown in Fig.~\ref{fig:map_param_cosmic_var}. 
Similarly, using the maps obtained in the ``Cosmic variance'' estimation, 
we compute the maps for $\Omega_{\kappa}$ and $q_{0}$. The difference maps $\Delta \Omega_{\kappa}$ and $\Delta q_{0}$, obtained by subtracting  $\Omega_{\kappa}^{\text{fid}}$ and $q_{0}^{\text{fid}}$,  respectively, from the corresponding maps, are shown Fig.~\ref{fig:map_param_cosmic_var}. Comparing the difference maps with the fiducial values, we measure fluctuations of order 10--100\% for $\Omega_{M},$ 1--30\% for $\Omega_{\Lambda}$, and  up to 3\% for $H_{0}.$ Similarly, we measure fluctuations of order 0.1--1000\% for $\Omega_{\kappa}$ and of order
10--60\% for $q_{0}.$ Although the fluctuations of $\Omega_{\kappa}$ are mostly small, they assume very high  values in the pixels where $\{\Omega_{M},\Omega_{\Lambda}\}$ fluctuate in the same direction. From the difference maps, we observe that the fluctuations in the $\Omega_{M}$ map are predominantly towards values higher than $\Omega_{M}^{\text{fid}},$ whereas the fluctuations in the $\Omega_{\Lambda}$ map are predominantly towards values lower than $\Omega_{\Lambda}^{\text{fid}}.$ The fluctuations in the $H_{0}$ map are equally distributed from lower to higher values than the fiducial. 

To compare our results with previous studies, we define the largest fluctuation in the difference maps as
\ba
\text{max}(\Delta \bar x_{ik})\equiv \text{max}(\bar x_{ik})-\text{min}(\bar x_{ik}),
\ea
which yields $\text{max}(\{\Delta \Omega_{M},\Delta \Omega_{\Lambda},\Delta H_0\})=\{0.352, 0.286, 3.84\},$
$\text{max}(\Delta \Omega_{\kappa})=0.363,$ and $\text{max}(\Delta q_{0})=0.452.$
We recall the results obtained in other SN studies. 
Kalus et al. found $2.0<\text{max}(\Delta H_{0})<3.4~\text{km~s}^{-1}\text{Mpc}^{-1}$ from the Union 2 data for fixed $q_{0}=-0.601$ \citep{kalus_2013}. Bengaly et al. found $\text{max}(\Delta H_{0})=\{4.6, 5.6\}~\text{km~s}^{-1}\text{Mpc}^{-1}$ and $\text{max}(\Delta q_{0})=\{2.56, 1.62\}$  from the Union 2.1 data and  JLA data, respectively \citep{bengaly_2015}. Our results for $H_{0}$ are intermediate between the previous results, whereas our results for $q_{0}$ are more stringent than those of Bengaly et al. Our results for $\{\Omega_{M}, \Omega_{\Lambda}\},$ and subsequently for $\Omega_{\kappa}$ are, to the best of our knowledge, the first measurements of this kind.

\begin{figure}[t] 
\centerline{
\includegraphics[width=\columnwidth]
{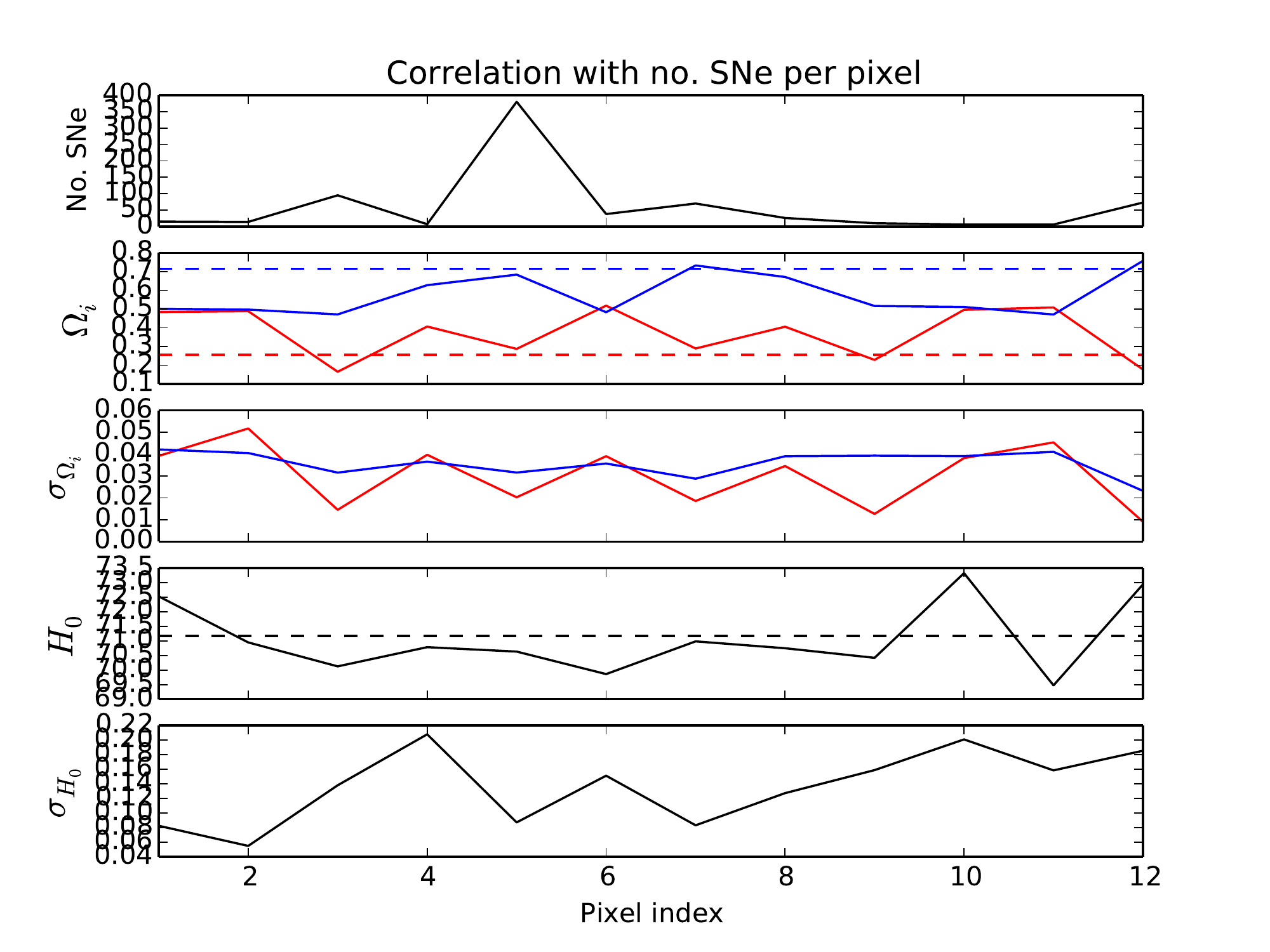}
}
\vspace{-.3cm}
\caption{
{\bf Correlation of the number of SNe per pixel with the fluctuations of the estimated parameters about the fiducial values.}  At each pixel we plot a) in the first panel, the number of SNe; b) in the second panel, the value of the parameters $\{\Omega_{M},\Omega_{\Lambda}\}$  as solid red and solid blue lines, respectively, with dashed red and dashed blue lines representing the corresponding fiducial values; c) in the third panel, the standard deviation $\{\sigma_{\Omega_{M}},\sigma_{\Omega_{\Lambda}}\}$ as solid red and solid blue lines, respectively; d) in the fourth plot, the value of $H_{0}$ as a solid black line, with the dashed black line representing the corresponding fiducial value; and e) in the fifth panel, the standard deviation $\sigma_{H_{0}}$ as a solid black line.
}
\label{fig:corr_per_pix}
\end{figure}

We also observe a mild negative correlation between the number of SNe in each pixel and the fluctuations about the fiducial values, which hints at a poor sampling derived from an inhomogeneous coverage of the sky, both in angular and redshift space, by the SN surveys.
This is further supported by the negative correlation between the number of SNe and the standard deviation of the parameters estimated in each pixel.
A poor sampling into pixels implies a poor sampling in redshift, which limits the constraint of the degeneracies among the parameters and hence compromises the estimation of the parameters.
Figure \ref{fig:corr_per_pix} illustrates these observations. 
In order to quantify this correlation, we define the correlation between the maps of  the parameters $\Delta \bar x_{ik}$ and the number of SNe per pixel $N_{\text{SN}k}$ by
\ba
&&\text{Corr}[\Delta \bar x_{ik},N_{\text{SN}k}]=\cr
&=&{ 
{\sum_k^{\text{Npixel}}
  (\Delta \bar x_{ik}-\left<\Delta \bar x_{ik}\right>_k)
  (N_{\text{SN}k}-\left<N_{\text{SN}k}\right>_k)}}\cr
&&\times{1 \over
\sqrt{
  \sum_k^{\text{Npixel}}(\Delta \bar x_{ik}-\left<\Delta \bar x_{ik}\right>_k)^2}}\cr
&&\times{1 \over
\sqrt{
  \sum_k^{\text{Npixel}}(N_{\text{SN}k}-\left<N_{\text{SN}k}\right>_k)^2}},
\ea
which takes values in $[-1,1].$
Similarly, we define the correlation between the sigma maps of the parameters  $\sigma_{\bar x_{ik}}$ and the number of SNe per pixel $N_{\text{SN}k}.$ We find 
$\text{Corr}[\Delta \bar x_{ik},N_{\text{SN}k}]=\{-0.394,0.397,-0.113\}$ 
and
$\text{Corr}[\sigma_{\bar x_{ik}},N_{\text{SN}k}]=\{-0.408,-0.452,-0.328\}.$
The values of $\text{Corr}[\Delta \bar x_{ik},N_{\text{SN}k}]$ show that the angular variations of $H_{0}$ are less sensitive to the number of SNe per pixel than the angular variations of $\Omega_{M}$ or $\Omega_{\Lambda}.$ 
The comparison of the values of $\text{Corr}[\Delta \bar x_{ik},N_{\text{SN}k}]$ with the values of $\text{Corr}[\sigma_{\bar x_{ik}},N_{\text{SN}k}]$ shows that the angular variations of the parameters are not as correlated with the number of SNe per pixel as the corresponding standard deviations.

The Galactic plane could contribute a spurious inhomogeneity in the local parameter estimation. However, an inspection of the pixels through which the Galactic plane crosses does not consistently show higher or lower values of the fluctuations to suggest a correlation.

These results are based on the assumption that both $\mu_{\text{data}}(z)$ and $\sigma_{\mu_{\text{data}}}$
are independent. The existence of correlations would effectively reduce the number of independent SNe, which would aggravate the parameter estimation in the less SN-populated pixels. Assuming independent
errors as a first approximation and having observed that the results are highly affected by the inhomogeneity of the SN surveys, the consideration of correlated errors would be a higher order effect and, hence, safely negligible for the current SN sample. 

To assess how the local estimation can be improved, we reduce the number of free parameters by fixing one parameter at a time to the corresponding fiducial value estimated from the complete sample. We observe that the estimation of the Hubble parameter is largely unaffected by the size of the pixel subsample. Conversely the estimation of the energy densities depends highly on the size of the pixel subsample. 

\section{Inhomogeneity tests}
\label{sec:test}

In order to further establish the validity of the method, 
we perform the following tests of the inhomogeneity of the SN surveys. In the first and second test, following \citet{kalus_2013}, we randomize the dependence of the redshift with the position by shuffling the SNe in $z.$ In the first test, we keep the inhomogeneous spatial distribution of the SNe (i.e. we keep the positions of the SNe as in the JLA sample) and, in the second test, we simulate a homogeneous spatial distribution (i.e. we choose random positions for the SNe). 
In the third test, we hypothesize a noise bias as a measure of the inhomogeneity in the SNe sampling and remove it from the local ``Cosmic variance'' estimation.

\begin{figure*}
\centerline{
\includegraphics[width=9cm]
{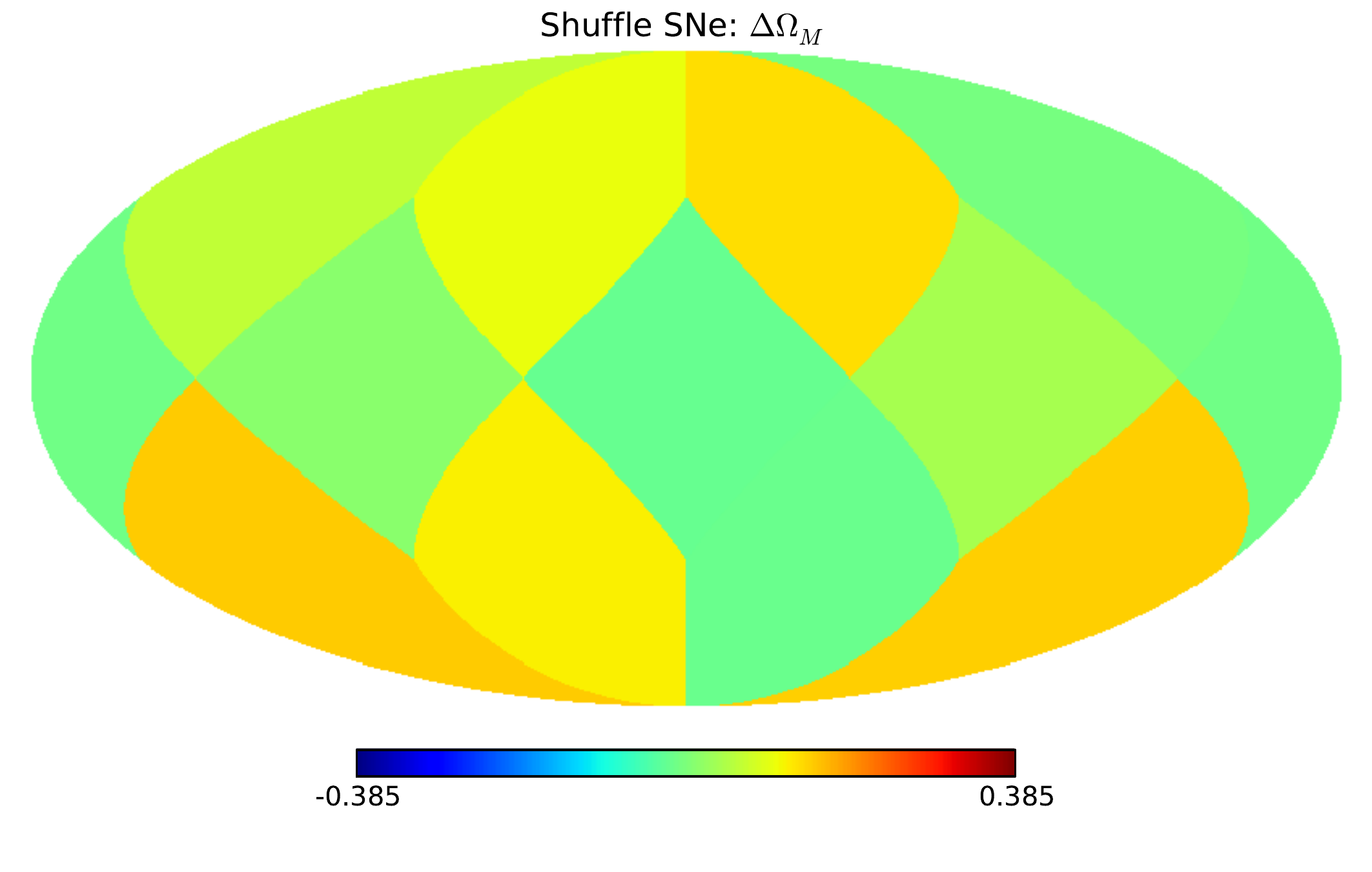}
\includegraphics[width=9cm]
{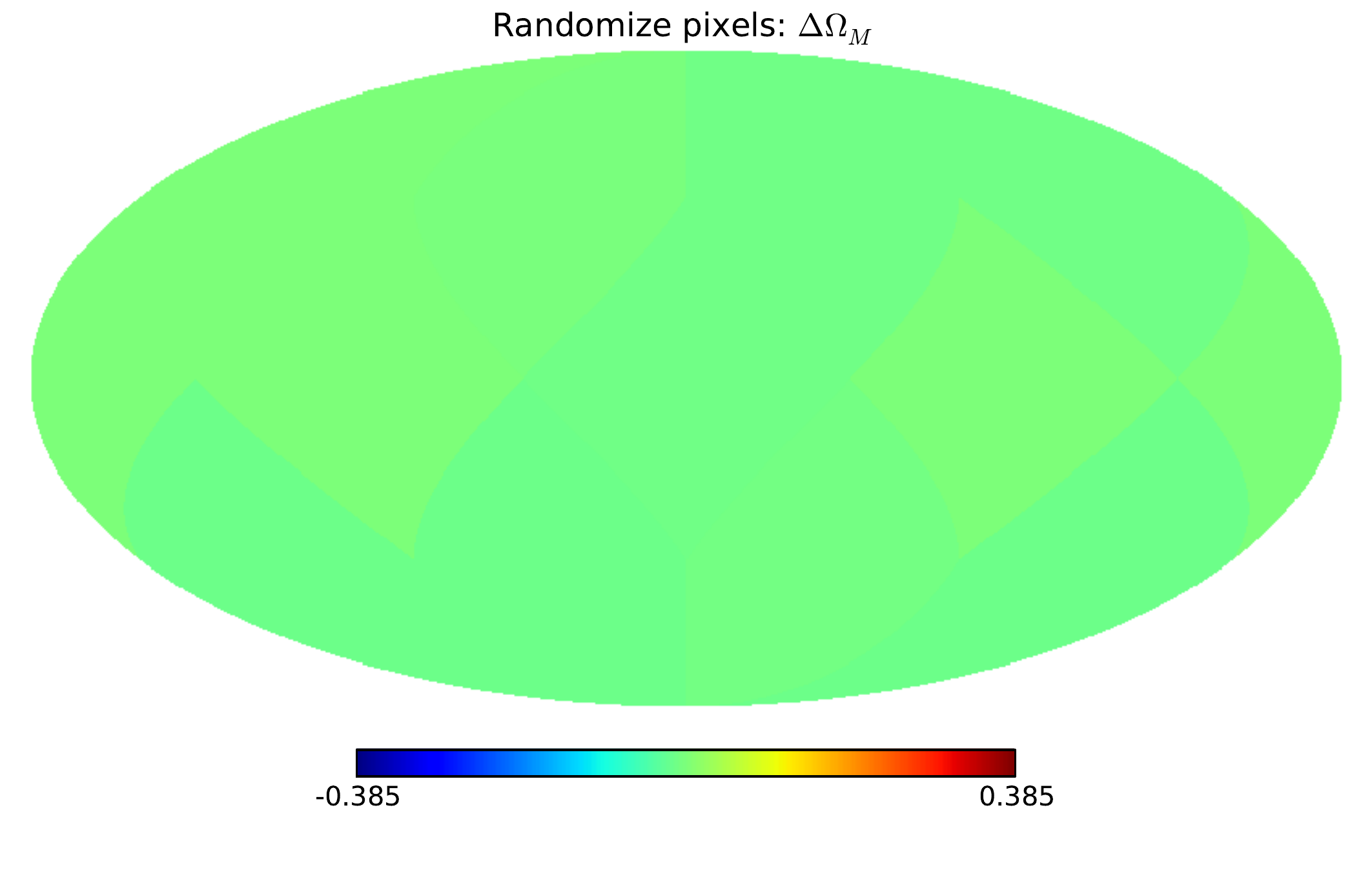}}
\centerline{
\includegraphics[width=9cm]
{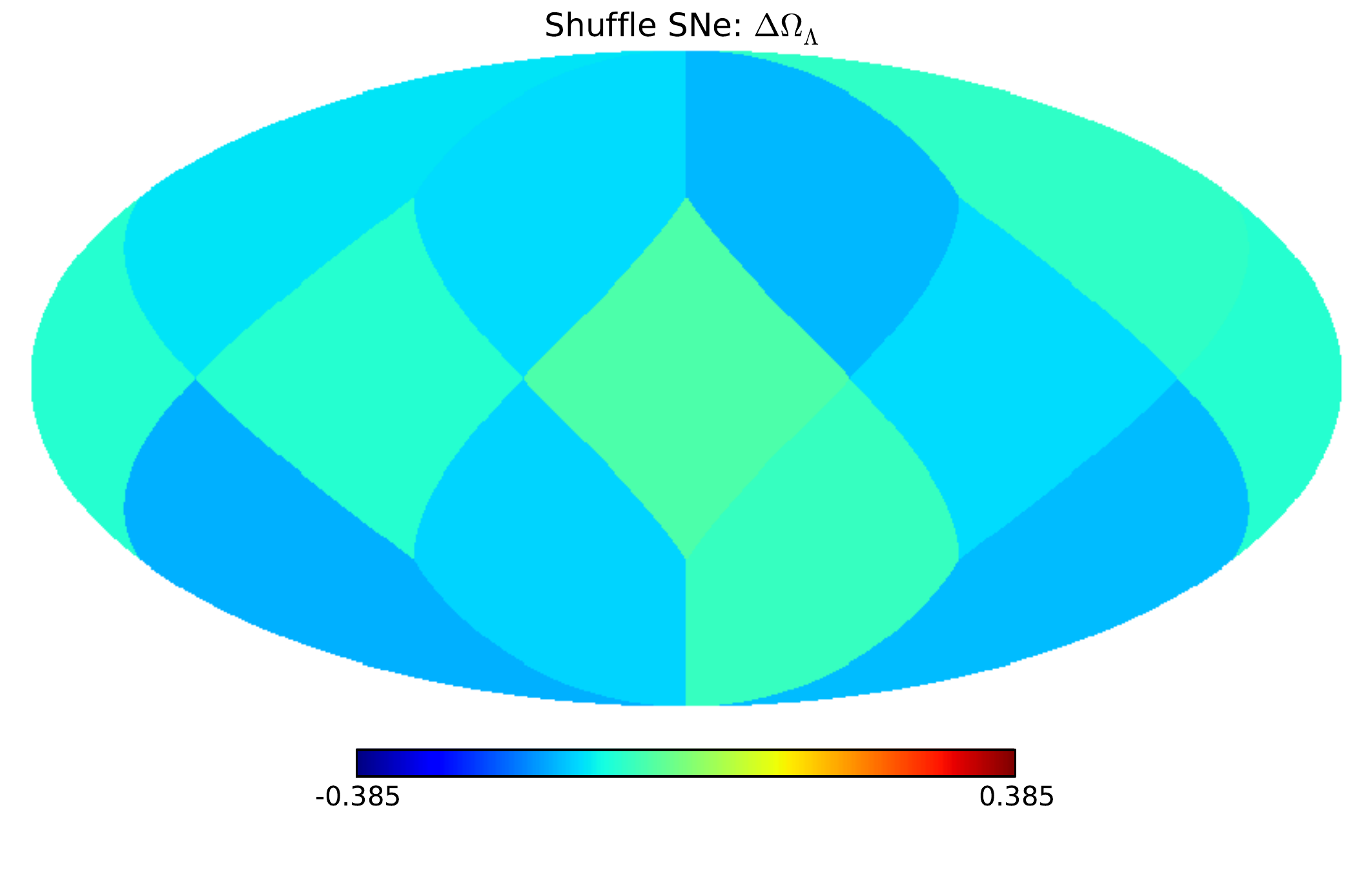}
\includegraphics[width=9cm]
{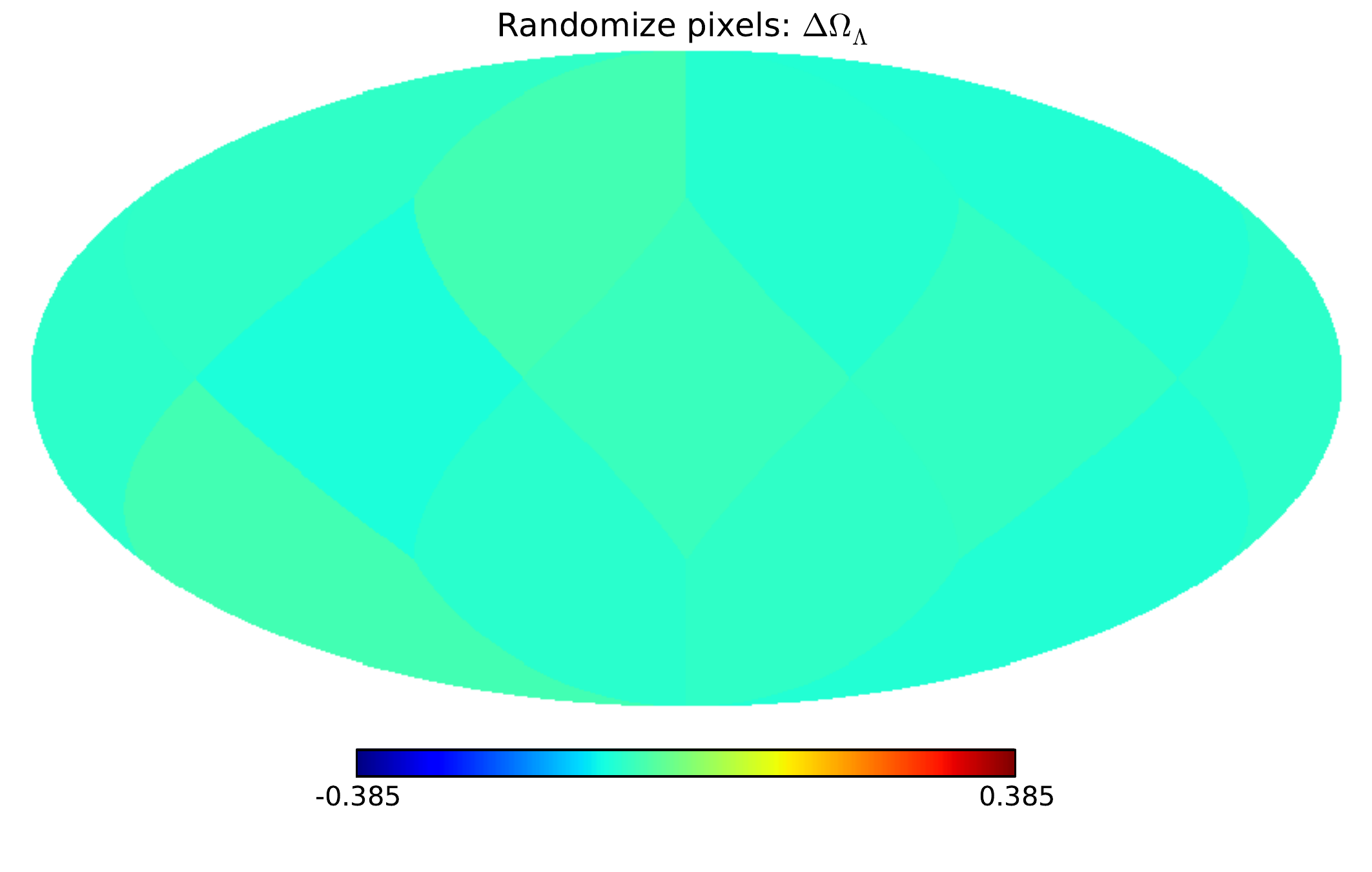}}
\centerline{
\includegraphics[width=9cm]
{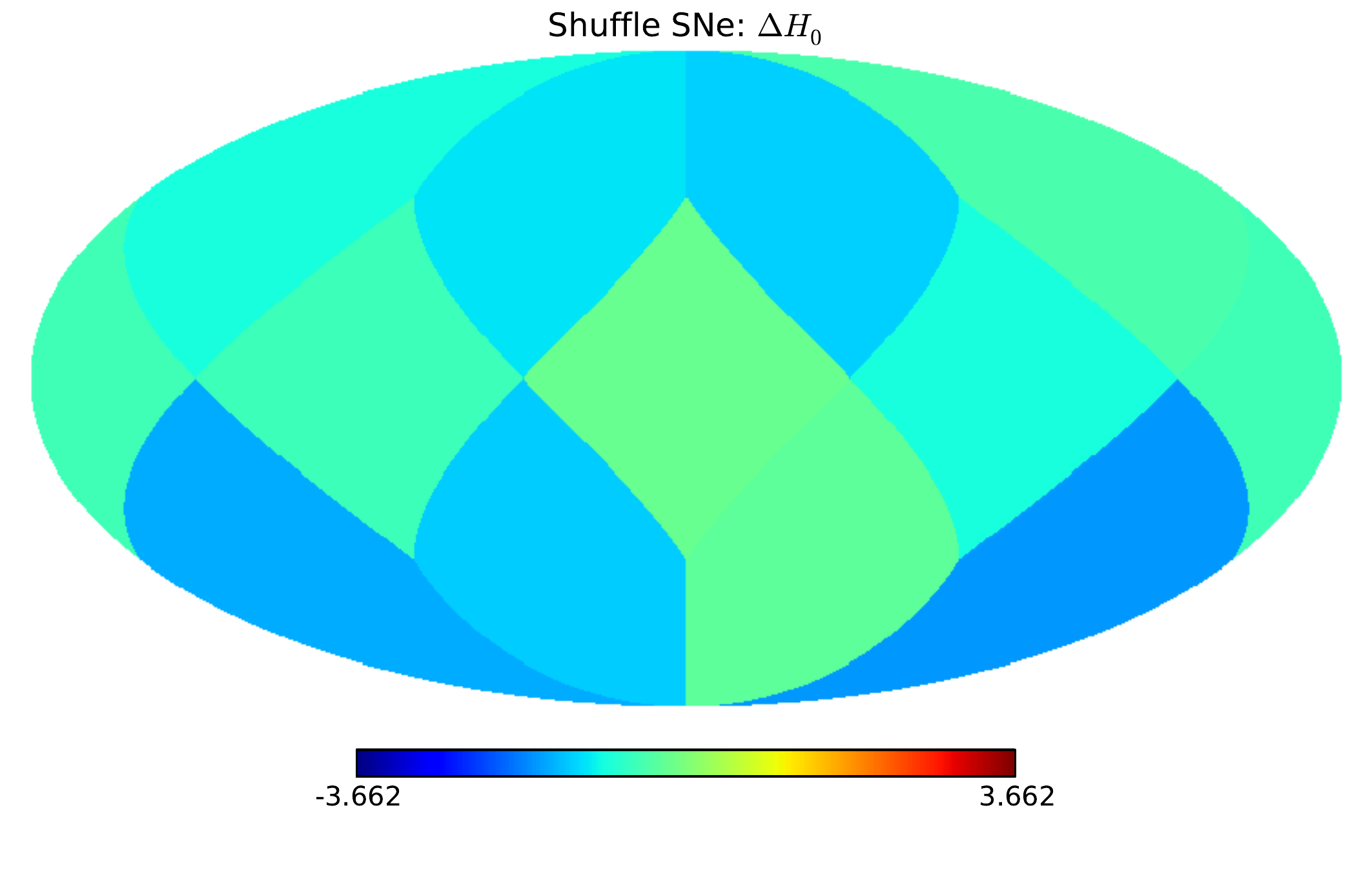}
\includegraphics[width=9cm]
{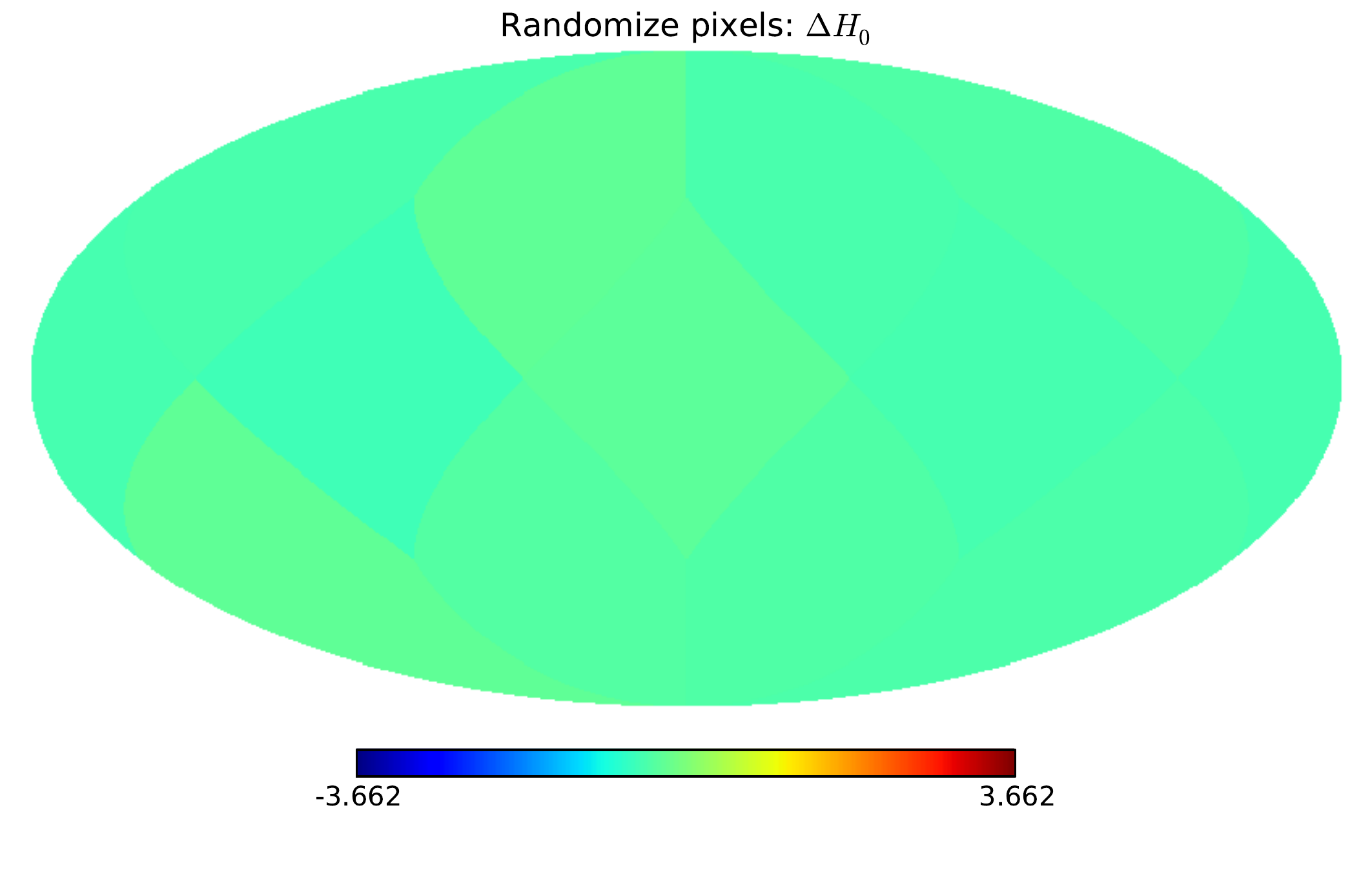}}
\vspace{-.3cm}
\caption{
{\bf Angular distribution of the parameters $\{\Omega_{M},\Omega_{\Lambda}, H_{0}\}_{k,\text{shuffle\_sn}}$  and $\{\Omega_{M},\Omega_{\Lambda}, H_{0}\}_{k,\text{random\_pix}}$ estimated at each pixel $k.$} The subsample of type Ia SNe in each pixel is used to estimate the value of the parameters in that pixel, which results in a map. Left column: test from shuffling the SNe. Right column: test from randomizing the pixelation. Top panel: difference map $\Delta\Omega_{M}=\Omega_{M}-\Omega_{M}^{\text{fid}}.$ Middle panel: difference map $\Delta \Omega_{\Lambda}=\Omega_{\Lambda}-\Omega_{\Lambda}^{\text{fid}}.$ Bottom panel: difference map $\Delta H_{0}=H_{0}-H_{0}^{\text{fid}}.$ The fiducial values are $\{\Omega_{M}^{\text{fid}},\Omega_{\Lambda}^{\text{fid}}, H_{0}^{\text{fid}}\}=\{0.256,   0.715,   71.17\}.$
}
\label{fig:map_param_test}
\end{figure*}

\subsection{Inhomogeneity test: Shuffle the SNe}

In this test, we randomize the dependence of redshift with the position for the original sample by shuffling the SNe in $z$ while keeping the positions in the sky. This way, we average out the inhomogeneities in the SN distribution in redshift for the same subsampling per pixel. 

Similar to the ``Cosmic variance'' estimation, we use the SN subsample in each pixel to estimate the maximum-likelihood values for the parameters $\{\Omega_{M},\Omega_{\Lambda}, H_{0}\}.$  We call this estimation ``Shuffle SNe.''
In each pixel, we ran 100 realizations of Markov chains of length $10^4.$ For each chain, the SN redshift is randomly shuffled for the same positions. In each pixel, the starting point of each realization is randomly generated as for the case of the complete sample. Also, as in the complete sample, we perform the burnt-in selection and thinning reduction of the Markov chains in each pixel. From each chain, the parameter estimation results in maps $x_{ijk}=\{\Omega_{M},\Omega_{\Lambda}, H_{0}\}_{jk,\text{shuffle\_sn}}.$ We average the maps for each parameter over the chains. Since the number of SNe in each pixel is the same for the different chains, we average the maps by taking the simple mean by pixel. We denote the resulting mean maps of the parameters by $\bar x_{ik}=\{\Omega_{M},\Omega_{\Lambda}, H_{0}\}_{k,\text{shuffle\_sn}}.$ 

For better visualization, we compute difference maps by subtracting the fiducial values $\{\Omega_{M}^{\text{fid}},\Omega_{\Lambda}^{\text{fid}}, H_{0}^{\text{fid}}\}$ from the corresponding maps.
Averaging over the various realizations of Markov chains at each pixel, we obtain the difference maps shown in  the left column of Fig.~\ref{fig:map_param_test}. For an easier comparison, we use the same colour scale for the maps that we used in the ``Cosmic variance'' estimation. 

Comparing the difference maps with $\{\Omega_{M}^{\text{fid}},\Omega_{\Lambda}^{\text{fid}}, H_{0}^{\text{fid}}\},$ we measure fluctuations of order 1--60\% for $\Omega_{M},$  1--20\% for $\Omega_{\Lambda}$, and  up to 3\% for $H_{0}.$ 
We observe that the fluctuations in the $\Omega_{M}$ are predominantly towards values higher than $\Omega_{M}^{\text{fid}},$ whereas the fluctuations in the $\Omega_{\Lambda}$ and $H_{0}$ are predominantly towards values lower than the corresponding fiducial values. 
We also observe the same negative correlation between the number of SNe in each pixel and the fluctuations about the fiducial values as in the ``Cosmic variance'' estimation. 

These fluctuations measure the inhomogeneity in the SN sampling, as observed in Fig.~\ref{fig:jla_sn}, which is reflected in the inhomogeneity of the subsampling into pixels. 
As observed, increasing the number of realizations increasingly homogenizes the sampling in each pixel and hence reduces the fluctuations about the fiducial values in pixels with large subsamples of SNe; however, increasing the number of realizations does not affect the estimation in pixels with small subsamples of SNe because the cosmic variance is large. For a sufficiently large number of realizations, such that the average is consistent with a homogeneous sampling into pixels with large subsamples of SNe, we hypothesize that this test measures the noise bias in the parameter estimation because of the inhomogeneity of the SN sampling. This noise bias is measured by the average maps produced in this test. Hence by removing this noise bias from the ``Cosmic variance'' estimation, we would obtain a parameter estimation that is independent of sampling inhomogeneities albeit with increased error.

\subsection{Inhomogeneity test: Randomize the pixelation}

In this test, we randomize the dependence of the redshift with the position for a homogeneous sample per pixel by distributing the SNe in randomly chosen positions in the sky. This way we are averaging out inhomogeneities in the SN distribution in redshift and  we simultaneously  average out inhomogeneities in the sampling. 

Similar to the ``Cosmic variance'' estimation, we use the SN subsample in each pixel to estimate the maximum-likelihood values for the parameters $\{\Omega_{M},\Omega_{\Lambda}, H_{0}\}.$ We call this estimation ``Random pixelation.''
In each pixel, we ran 100 realizations of Markov chains of length $10^4.$ For each chain, the SNe redshift is randomly distributed in a homogeneous way throughout the sky. In each pixel, the starting point of each realization is randomly generated as for the case of the complete sample. Also, as in the complete sample, we perform the burnt-in selection and thinning reduction of the Markov chains in each pixel. From each chain, the parameter estimation results in maps  $x_{ijk}=\{\Omega_{M},\Omega_{\Lambda}, H_{0}\}_{jk,\text{random\_pix}}.$ We average the maps for each parameter over the chains. Since the number of SNe in each pixel varies among the chains, we average the maps by taking the mean weighted by the number of SNe in each pixel. We denote the resulting mean maps of the parameters by $\bar x_{ik}=\{\Omega_{M},\Omega_{\Lambda}, H_{0}\}_{k,\text{random\_pix}}.$ 

For better visualization, we compute difference maps by subtracting the fiducial values $\{\Omega_{M}^{\text{fid}},\Omega_{\Lambda}^{\text{fid}}, H_{0}^{\text{fid}}\}$ off the corresponding maps. Averaging over the various realizations of Markov chains at each pixel, we obtain the difference maps $\{\Delta\Omega_{M},\Delta\Omega_{\Lambda}, \Delta H_{0}\}_{k,\text{random\_pix}}$ shown in the right column of Fig.~\ref{fig:map_param_test}. 

Comparing the difference maps with $\{\Omega_{M}^{\text{fid}},\Omega_{\Lambda}^{\text{fid}}, H_{0}^{\text{fid}}\},$ we measure fluctuations of order 10\% 
for $\Omega_{M}$ and $\Omega_{\Lambda},$ and of order less than 1\% for $H_{0}.$ These fluctuations measure the limitation of an equal subsampling per pixel of the original sample. Since these fluctuations are smaller than the error of the global parameter estimation, they are consistent with a homogeneous sampling in all pixels, which implies that the original sample is large enough to be divided into subsamples. 

\begin{figure*}
\centerline{
\includegraphics[width=9cm]
{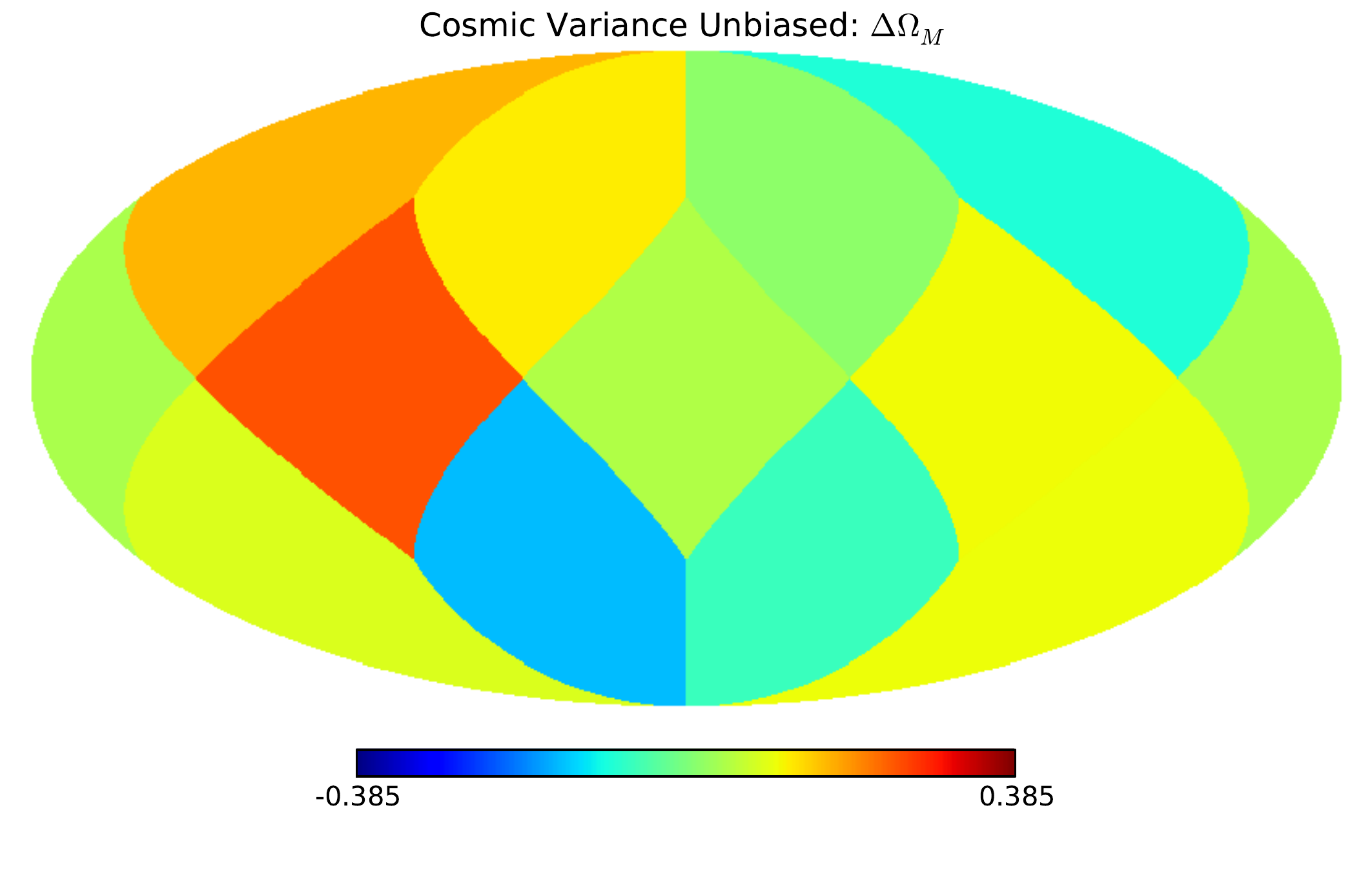}
\includegraphics[width=9cm]
{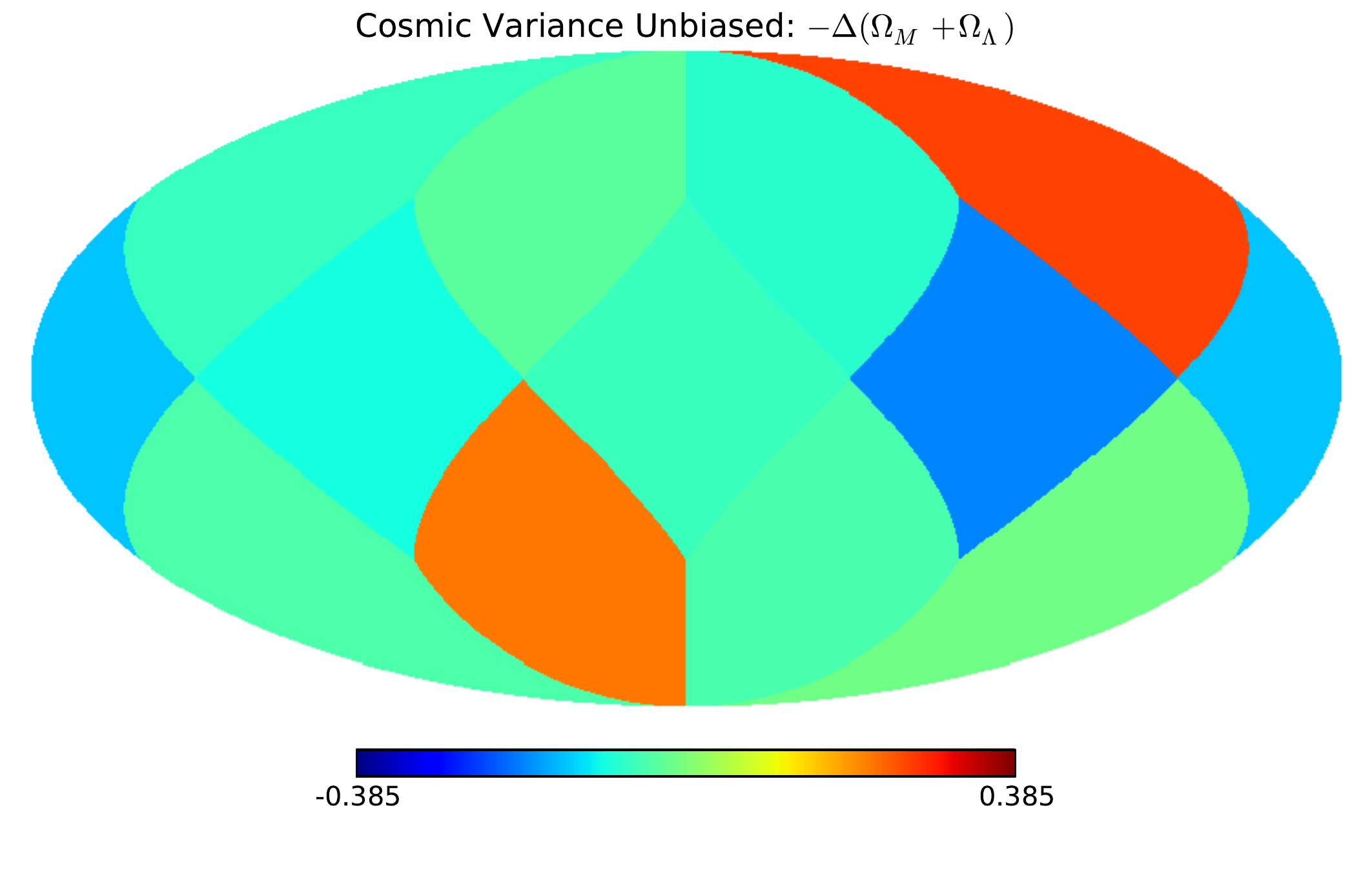}
}
\centerline{
\includegraphics[width=9cm]
{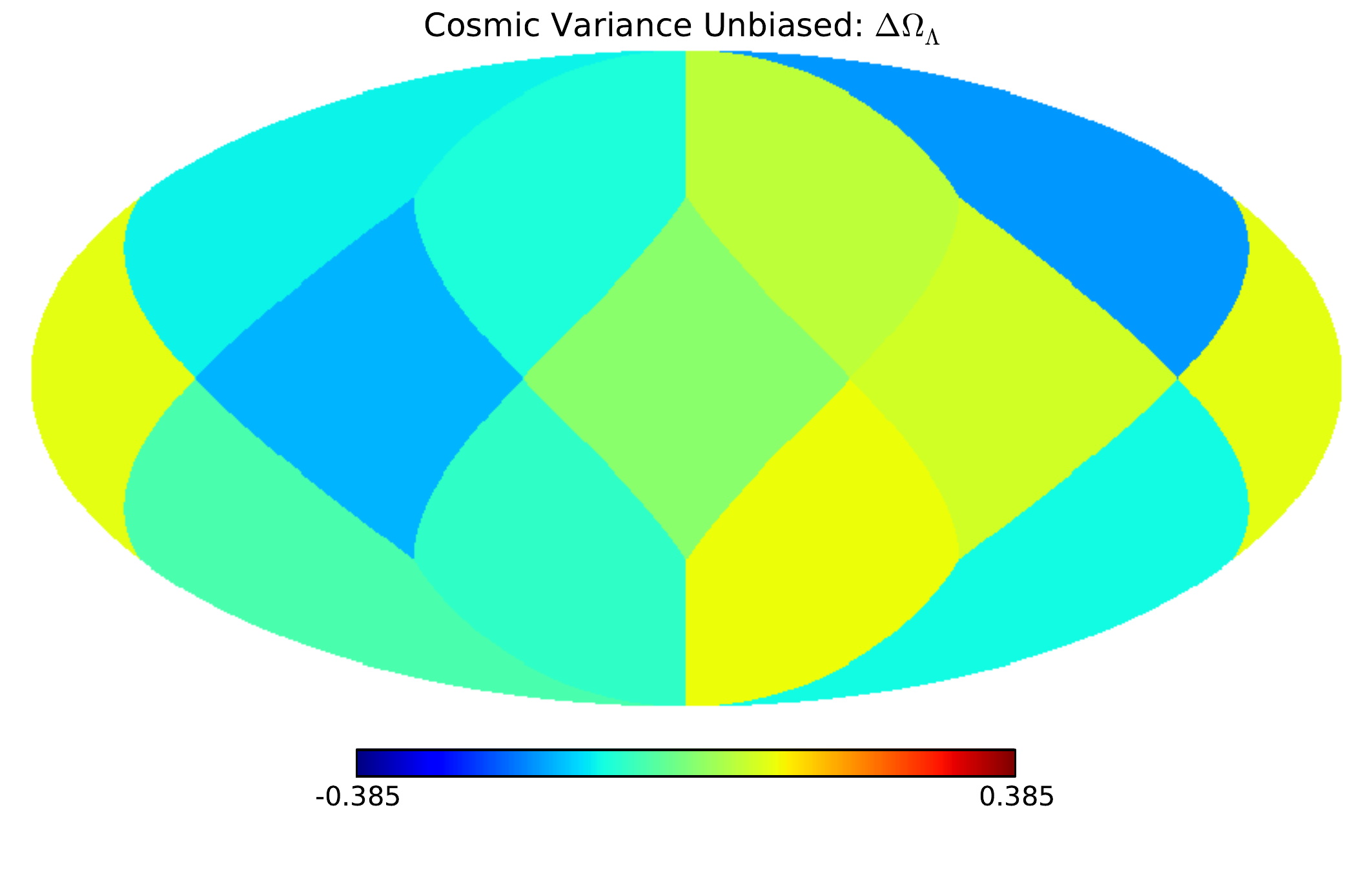}
\includegraphics[width=9cm]
{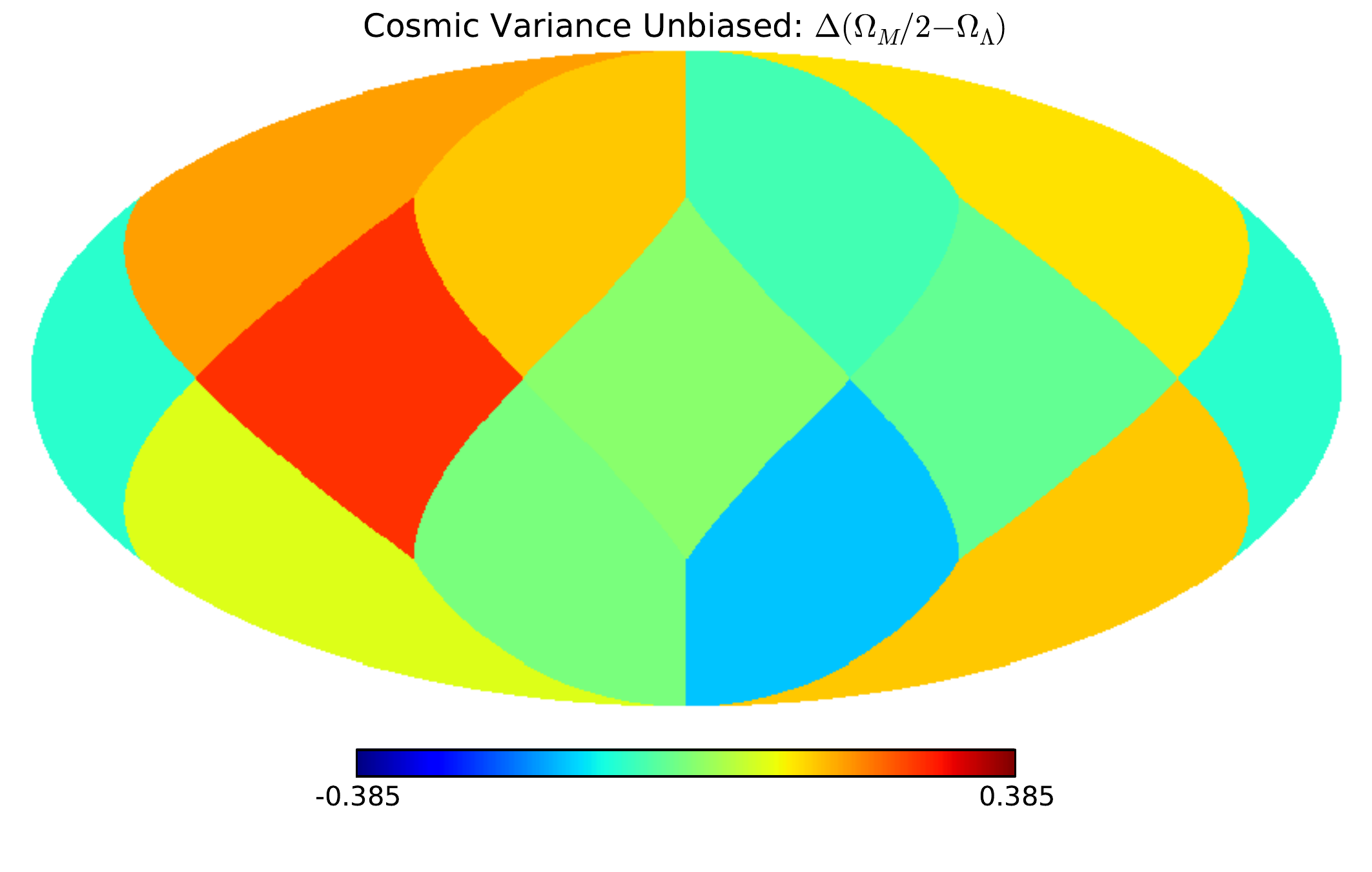}
}
\centerline{
\includegraphics[width=9cm]
{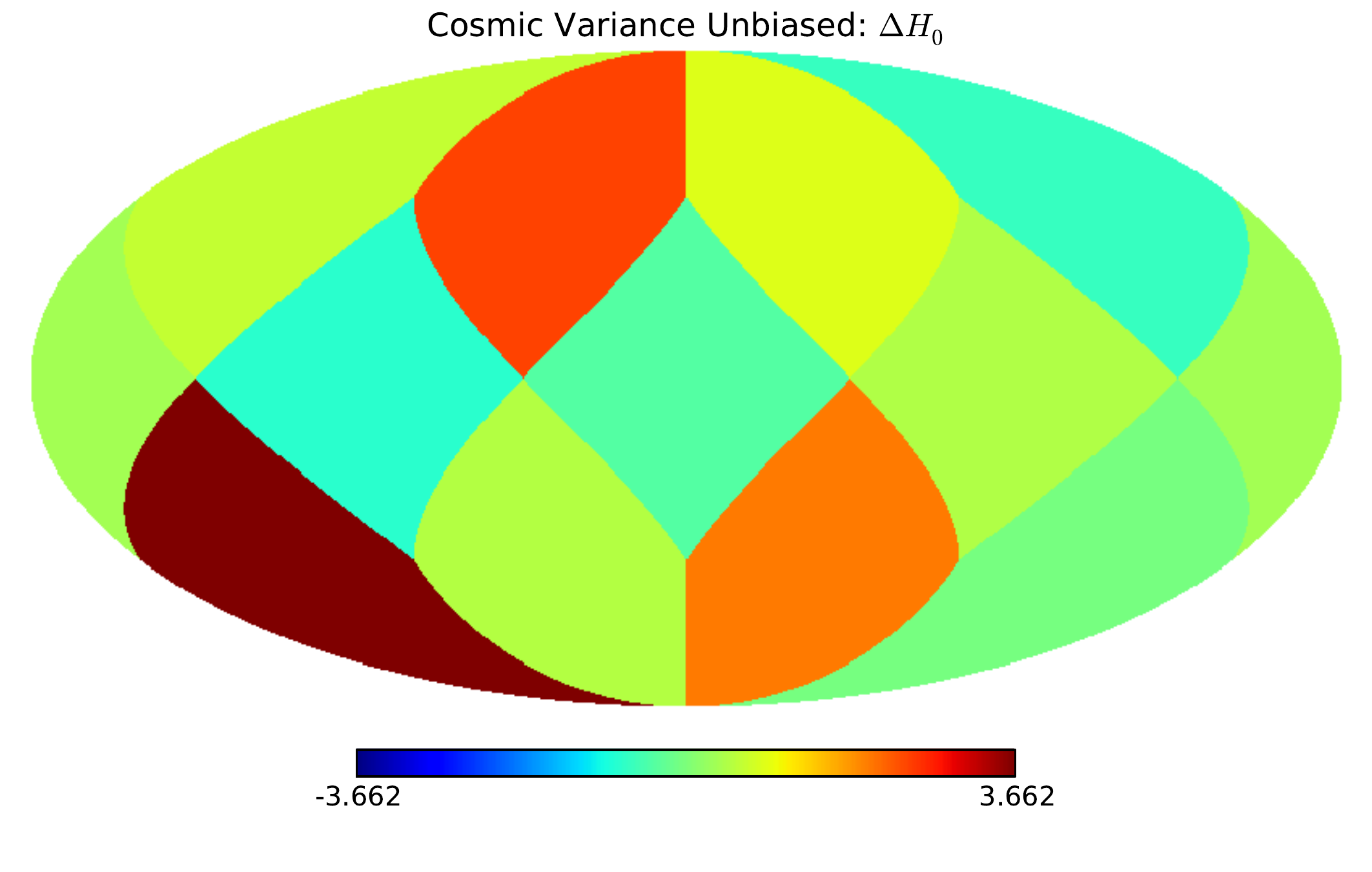}
}
\vspace{-.3cm}
\caption{
{\bf Angular distribution of the estimated parameters $\{\Omega_{M},\Omega_{\Lambda}, H_{0}\}_{k,\text{cosmic\_var}}$ at each pixel $k$ after  noise bias removal.} The subsample of type Ia SNe in each pixel is used to estimate the value of the parameters $\{\Omega_{M},\Omega_{\Lambda}, H_{0}\}$ in that pixel, resulting in a map for each parameter. Subtracting the maps $\{\Omega_{M},\Omega_{\Lambda}, H_{0}\}_{k,\text{shuffle\_sn}}$ averaged over the different realizations, results in the corresponding unbiased maps. Operating the estimated maps results in the derived maps for $\Omega_{\kappa}$ and $q_{0}.$ Subtracting the fiducial values, results in the corresponding difference maps. 
The fiducial values are $\{\Omega_{M}^{\text{fid}},\Omega_{\Lambda}^{\text{fid}}, H_{0}^{\text{fid}}\}=\{0.256,   0.715,   71.17\}.$
}
\label{fig:map_param_unbias_cosmic_var}
\end{figure*}

\subsection{Inhomogeneity test: Remove the noise bias}
We now proceed to test the hypothesis suggested in the context of the ``Shuffle SNe'' estimation. We let $x_{ijk}$ and $x_{ij^{\prime}k}^{\text{bias}}$ denote maps produced by $j$ chains of the ``Cosmic variance'' estimation and $j^{\prime}$ chains of the ``Shuffle SNe'' estimation, respectively. Then for each chain $j,$ we define the unbiased maps as $x_{ijk}^{\text{unbias}}\equiv x_{ijk}-\bar x_{ik}^{\text{bias}}+x_i^{\text{fid}},$ where $\bar x_{ik}^{\text{bias}}\equiv\big< x_{ij^{\prime}k}^{\text{bias}}\big>_{j^{\prime}}$ is the average over the $j^{\prime}$ realizations of the ``Shuffle SNe'' estimation. 
The unbiased maps averaged over the chains $j$ are $\bar x_{ik}^{\text{unbias}}\equiv \big<x_{ijk}\big>_{j}-\bar x_{ik}^{\text{bias}}+x_i^{\text{fid}},$ and the corresponding unbiased difference maps are $\Delta \bar x_{ik}^{\text{unbias}}\equiv \bar x_{ik}^{\text{unbias}}-x_{i}^{\text{fid}}.$ 
We show $\Delta \bar x_{ik}^{\text{unbias}}=\left\{\Delta \Omega_{M},\Delta \Omega_{\Lambda}, \Delta H_{0}\right\}^{\text{unbias}}_{k}$ in Fig.~\ref{fig:map_param_unbias_cosmic_var} to be compared with Fig.~\ref{fig:map_param_cosmic_var}. 
Comparing the unbiased difference maps with $\{\Omega_{M}^{\text{fid}},\Omega_{\Lambda}^{\text{fid}}, H_{0}^{\text{fid}}\},$ we measure fluctuations of order 5--95\% for $\Omega_{M},$ 1--25\% for $\Omega_{\Lambda}$, and up to 5\% for $H_{0}.$ Similarly, we measure fluctuations of order 35--900\% for $\Omega_{\kappa}$ and  1--50\% for $q_{0}.$ 

\begin{figure}[t] 
\centerline{
\includegraphics[width=\columnwidth]
{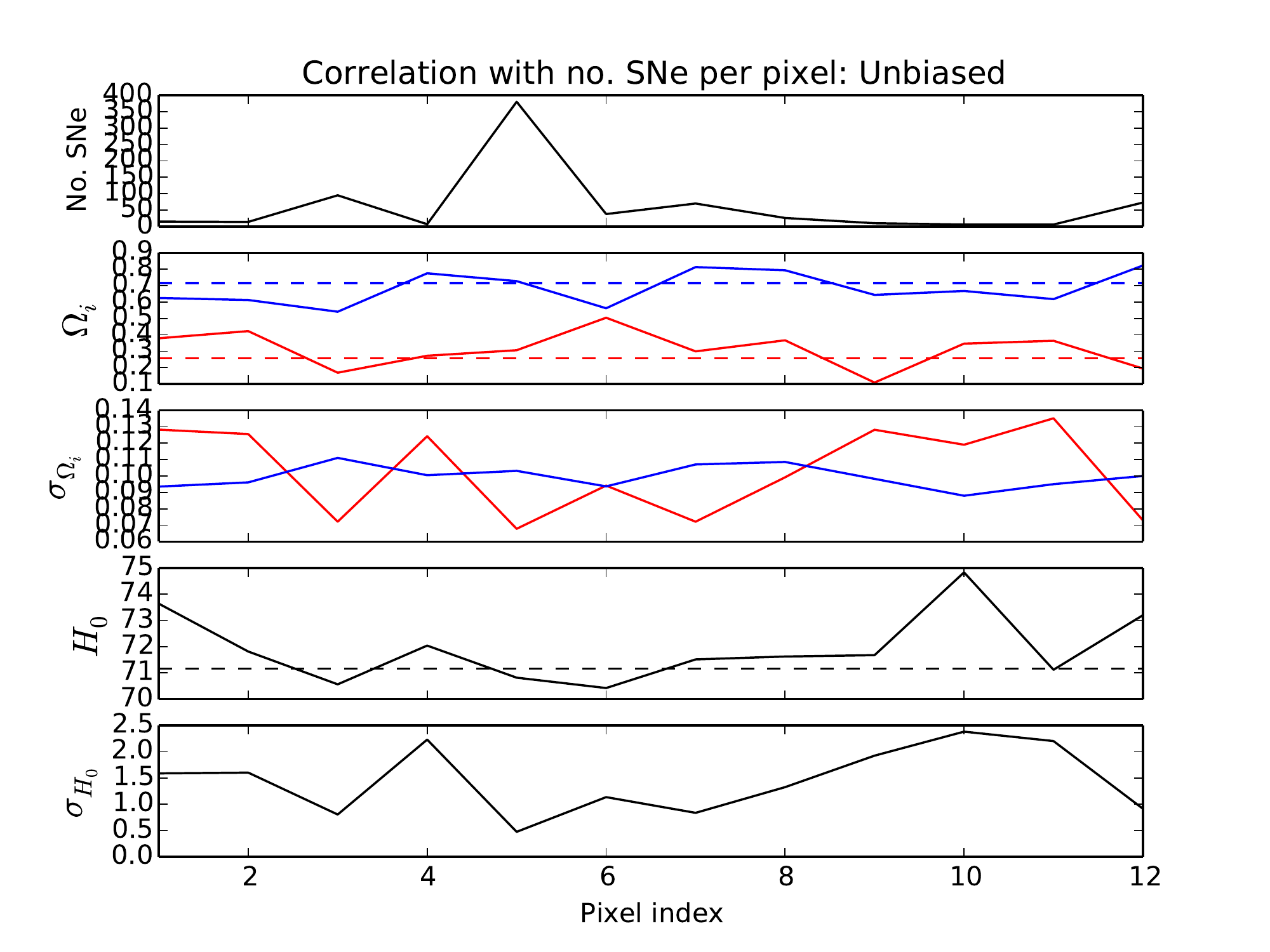}}
\vspace{-.3cm}
\caption{
{\bf Correlation of the number of SNe per pixel with the estimated parameter fluctuations about the fiducial values after the noise bias removal.} At each pixel we plot a) in the first panel, the number of SNe; b) in the second panel, the value of the parameters $\{\Omega_{M},\Omega_{\Lambda}\}$  as solid red and solid blue lines, respectively, with the dashed red and dashed blue lines representing the corresponding fiducial values; c) in the third panel, the standard deviation $\{\sigma_{\Omega_{M}},\sigma_{\Omega_{\Lambda}}\}$ as solid red and solid blue lines, respectively; d) in the fourth plot, the value of $H_{0}$ as a solid black line, with the dashed black line representing the corresponding fiducial value; e) in the fifth panel, the standard deviation $\sigma_{H_{0}}$ as a solid black line.
}
\label{fig:corr_per_pix_unbias}
\end{figure}

For the correlations of the number of SNe per pixels with the difference maps and sigma maps, we find $\text{Corr}[\Delta \bar x_{ik}^{\text{unbias}},N_{\text{SN}k}]=\{-0.117,0.162,-0.343\}$ 
and 
$\text{Corr}[\sigma_{\bar x_{ik}^{\text{unbias}}},N_{\text{SN}k}]=\{-0.653,-0.348,-0.689\}$, respectively. For the unbiased maps in comparison with the original maps, the correlation with the difference maps decreases, whereas the correlation with the error maps increases, as expected. 
These results seem to indicate the validity of $\bar x_{ik}^{\text{bias}}$ as a measure of the noise bias from the inhomogeneity of the SN sampling.
Figure \ref{fig:corr_per_pix_unbias} illustrates these observations to be compared with Fig.~\ref{fig:corr_per_pix}.

\begin{figure*}
\centerline{
\includegraphics[width=10cm]
{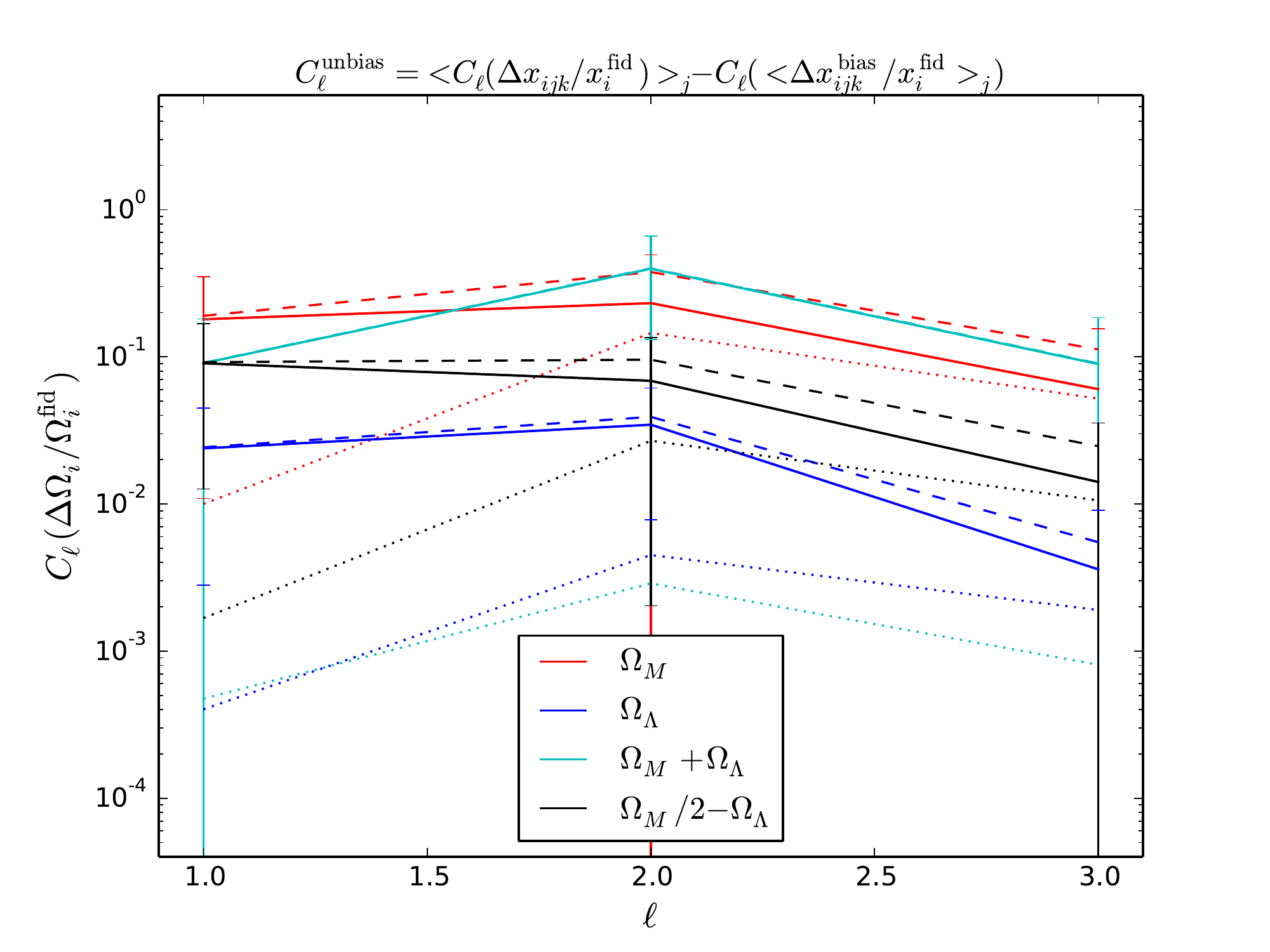}
\includegraphics[width=10cm]
{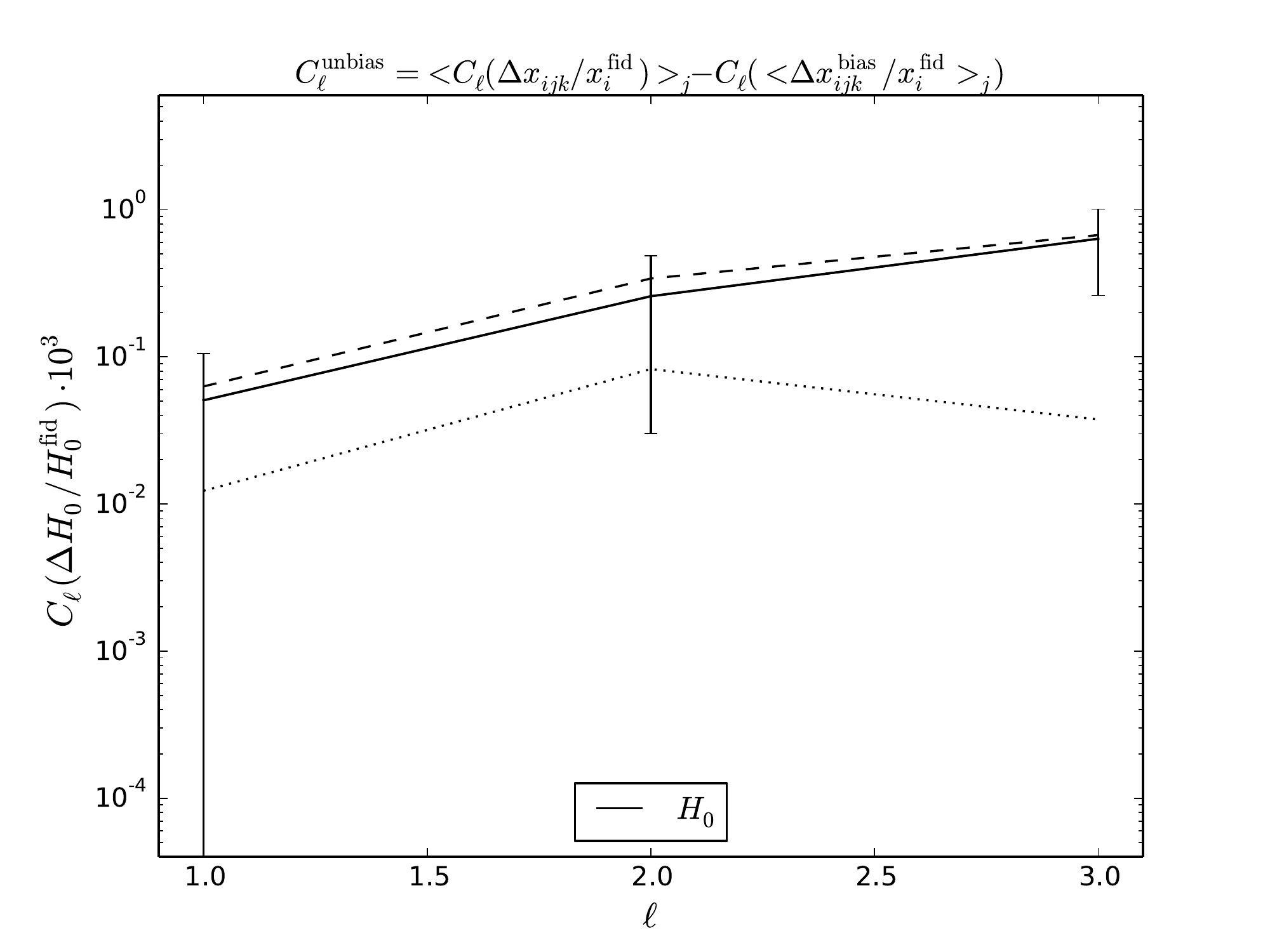}
}
\vspace{-0.3cm}
\caption{
{\bf Power spectrum of the parameters estimated from the JLA type Ia SN sample.}
The dashed lines represent the average power spectra from the cosmic variance estimation, the dotted lines represent the power spectra of the average shuffle SNe estimation, and the solid lines are the unbiased power spectra defined as the difference between the former two. Left panel: power spectra of the matter energy density $\Omega_{M}$ and dark energy density $\Omega_{\Lambda}$  
(estimated parameters), curvature energy density $\Omega_{\kappa}$, and the deceleration parameter $q_{0}$ (derived parameters).
Right panel: power spectrum of the Hubble parameter $H_0$ (estimated parameter).
}
\label{fig:ps_param}
\end{figure*}

\section{Angular power spectrum of the parameters}
\label{sec:power_spec}

In order to probe anisotropies in the parameters, we compute the angular power spectrum of the difference maps that result from each realization normalized to the corresponding fiducial parameter.  We consider the cosmic variance estimation. For each estimated parameter $i,$ for each realization $j,$ we define the map of the fluctuations about the fiducial value at each pixel $k$ as $\delta x_{ijk}=(x_{ijk}-x_{i}^{\text{fid}})/x_{i}^{\text{fid}}.$ The map defined over the full sky can be decomposed into spherical harmonics according to
\ba
\delta x_{ijk}(\boldsymbol{n}_{k})=\sum_{\ell=0}^{\infty}\sum_{m=-\ell}^{\ell}a_{\ell m,ij}Y_{\ell m}(\boldsymbol{n}_{k}),
\ea
where each pixel $k$ defines a unit direction vector $\boldsymbol{n}_{k}$ on the sky.
The harmonic coefficients are given by
\ba
a_{\ell m,ij}=\int d\boldsymbol{n}_{k}~\delta x_{ijk}(\boldsymbol{n}_{k})Y_{\ell m}^{\ast}(\boldsymbol{n}_{k}),
\ea
which for a pixelated map with $N_{\text{pix}}$ pixels can be estimated  by
\ba
\hat a_{\ell m,ij}={4\pi\over N_{\text{pixel}}}\sum_{k=1}^{N_{\text{pixel}}}\delta x_{ijk}Y_{\ell m}^{\ast}(\boldsymbol{n}_{k}).
\ea 
Assuming statistical isotropy, the $\hat a_{\ell m,ij}$ can be used to estimate the power spectrum $\hat C_{\ell,ij}$ according to
\ba
\big<a_{\ell m,ij}a_{\ell^{\prime}m^{\prime}}^{\ast}\big>=\delta_{\ell\ell^{\prime}}\delta_{mm^{\prime}}C_{\ell,ij}, 
\ea
which can be inverted to yield the power spectrum of the parameter $i$ from the map realization $j$
\ba
\hat C_{\ell,ij}={1\over (2\ell+1)}\sum_{m=-\ell}^{\ell}\vert \hat a_{\ell m,ij}\vert^2.
\ea
The variable $\ell$ can be written approximately as  $\ell=2\pi/\theta,$ where $\theta$ ranges from the angular size of the map to twice the size of the pixel, which corresponds to the Nyquist frequency. The minimum value of $\ell$ is determined by the largest scale probed by the map, hence $\theta=2\pi$ and $\ell_{\text{min}}=1.$ Conversely the maximum value of $\ell$ is determined by the smallest scale probed by the map, hence $\theta=2\times \text{(pixel size)}$ and $\ell_{\text{max}}=3.$ 

Averaging the power spectra over the different realizations, we obtain the mean power spectrum $\hat C_{\ell,i}$ for the parameters $\{\Omega_{M},\Omega_{\Lambda}, H_{0}\}_{\text{cosmic\_var}}$ 
\ba
\hat C_{\ell,i}=\left<\hat C_{\ell,ij}\right>_{j}.
\ea
The variance of the estimator 
is given by
\ba
\text{Var}[{\hat C_{\ell,i}}]^{\text{estimator}}={2\over {2\ell+1}}\hat C_{\ell,i}^2.
\ea
Hence the total variance of $\hat C_{\ell,i}$ is the sum of the sample variance about the mean value and the variance of the estimator 
\ba
\text{Var}[{\hat C_{\ell,i}}]=\text{Var}[{\hat C_{\ell,i}}]^{\text{sample}}+\text{Var}[{\hat C_{\ell,i}}]^{\text{estimator}}.
\ea
The resulting power spectra are plotted in Fig.~\ref{fig:ps_param} as dashed lines. 
 
For the density parameters, as well as for the deceleration parameter, we observe that the power spectrum has a maximum at the quadrupole ($\ell=2).$
The $\Omega_M$ power spectrum is significantly larger than the $\Omega_{\Lambda}$ power spectrum over all scales, ranging from 5 to 20 times larger.
For the Hubble parameter, the power spectrum always increases with the multipole.
The quadrupole is approximately $\{2.0, 1.6, 5.4\}$ times the dipole ($\ell=1$) 
in the $\{\Omega_{M},\Omega_{\Lambda}, H_{0}\}_{\text{cosmic\_var}}$ power spectra, respectively. 

In \citet{bengaly_2015}, the dipole was the dominant scale of the $H_0$ power spectrum produced from the low-redshift ($z\le 0.20$) SNe of the JLA sample. This effect was attributed either to the bulk motion (i.e. the recessional velocity component due to the global expansion) or to the inhomogeneous distribution of the data set.

In order to establish the validity of the method for the power spectrum analysis, we use the average ``Shuffle SNe'' estimation over $j^{\prime}$ realizations, denoted by $\bar x_{ik}^{\text{bias}},$ as a measure of the noise bias due to the inhomogeneity of the SN sampling. For each estimated parameter $i,$ we define the average map of the fluctuations about the fiducial value at each pixel $k$ as $\delta \bar x_{ik}^{\text{bias}}=(\bar x_{ik}^{\text{bias}}-x_{i}^{\text{fid}})/x_{i}^{\text{fid}}.$
From the decomposition into spherical harmonics of $\delta \bar x_{ik}^{\text{bias}},$ we compute $\hat C_{\ell,i}^{\text{bias}}.$ We define the unbiased power spectrum as 
\ba
\hat C_{\ell,i}^{\text{unbias}} =\hat C_{\ell,i}- \hat C_{\ell,i}^{\text{bias}} 
\ea
with total variance 
\ba
\text{Var}[{\hat C_{\ell,i}^{\text{unbias}}}]=\text{Var}[{\hat C_{\ell,i}}]
+\text{Var}[{\hat C_{\ell,i}}^{\text{bias}}]^{\text{estimator}}.
\ea
The resulting power spectra are plotted in Fig.~\ref{fig:ps_param} with the bias power spectra as dotted lines and the unbiased power spectra as solid lines. 
The unbiased power spectra of the density parameters and of the Hubble parameter have the same behaviour, albeit less pronounced, as the corresponding power spectra before the noise bias removal. The exception is the deceleration parameter whose subtle maximum at $\ell=2$ is erased with the noise bias removal. After the noise bias removal, 
the quadrupole is approximately $\{1.3, 1.4, 5.1\}$ times the dipole 
in the $\{\Omega_{M},\Omega_{\Lambda}, H_{0}\}_{\text{cosmic\_var}}$ power spectra, respectively. 
Although these results seem to indicate the validity of $\hat C_{\ell,i}^{\text{bias}}$ as a measure of the noise bias and the dominance of the quadrupole as the presence an inhomogeneity scale, the errors are too large to validate the measurements, thus rendering the results compatible with a flat spectrum.

\section{Average over the local estimations}
\label{sec:backreact}

We proceed to compute average values of the parameters from the corresponding locally estimated values. We use the maps obtained in the ``Cosmic variance'' estimation. Given the mean maps over the chains $\bar x_{ik}$, we want to average the map of each parameter over the pixel subsamples. 

One way of averaging consists in taking the mean weighted by the inverse of the variance $w_k=1/\text{Var}[\bar x_{ik}]$ of the parameter in each pixel $k,$ 
\ba
\bar x_i=\big< \bar x_{ik}\big>_{k}={\sum_{k}^{N_{\text{pixel}}}w_k~\bar x_{ik} \over {\sum_{k}^{N_{\text{pixel}}}w_k}}.
\ea
For the estimated parameters, we find
$\{\overline\Omega_{M},\overline \Omega_{\Lambda},\overline H_{0}\}
=\{0.242,0.607,71.06\}\pm \{0.099,0.115,0.87\}$
and 
$\{\overline\Omega_{M},\overline \Omega_{\Lambda},\overline H_{0}\}_{\text{flat}}
=\{0.276,0.724,70.93\}\pm \{0.041,0.041,0.64\}.$
These values are consistent with, but systematically smaller than, the fiducial values estimated from the complete sample (see Table \ref{table:param_pix}.)
After the noise bias removal, we find 
$\{\overline\Omega_{M},\overline \Omega_{\Lambda},\overline H_{0}\}_{\text{unbias}}
=\{0.294,0.679,71.37\}\pm \{0.115,0.120,0.94\}$
and 
$\{\overline\Omega_{M},\overline \Omega_{\Lambda},\overline H_{0}\}_{\text{unbias,flat}}
=\{0.256,0.715,71.17\}\pm \{0.056,0.056,0.90\}.$ 
These values are consistent with the fiducial values. We also observe that the flatness condition yields a lower $H_{0}$ than in the absence of the flatness condition, which is the opposite trend of that observed in the estimation from the complete sample.

Similarly, we compute the pixel average of the maps for $\Omega_{\kappa}$ and $q_{0}$ derived from the maps obtained in the ``Cosmic variance'' estimation. The average value of $\Omega_{\kappa}$ over the pixels is 
$\overline\Omega_{\kappa}=0.078\pm0.131$ and $\overline\Omega_{\kappa,\text{flat}}=0.000\pm0.058,$ and  
after the noise bias removal it is
$\overline\Omega_{\kappa,\text{unbias}}=0.002\pm0.135$ and 
$\overline\Omega_{\kappa,\text{unbias,flat}}=0.029\pm0.079.$ 
The average value of $q_{0}$ over the pixels is 
$\bar q_{0}=-0.451\pm 0.159$ and $\bar q_{0,\text{flat}}=-0.586\pm 0.061,$ 
and after the noise bias removal it is
$\bar q_{0,\text{unbias}}=-0.526\pm 0.172$ and $\bar q_{0,\text{unbias,flat}}=-0.587\pm 0.083$ 
(see Table \ref{table:param_pix}). 
After the noise bias removal, the pixel average values of both $\Omega_{\kappa}$ and $q_{0}$ become closer to the corresponding values estimated from the complete sample. 
We also observe that the flatness condition yields a smaller $q_{0}.$ 

Since we are averaging over inhomogeneous maps, 
we consider a toy model of an inhomogeneous space-time 
consisting of disjoint, locally homogeneous regions with different expansion factors. 
Hence another way of averaging consists in taking the mean weighted by the 3--volume $V_k$ of each pixel $k,$  
\ba
\bar x_i=\big< \bar x_{ik}\big>_{V_k}
={\sum_{k}^{N_{\text{pixel}}}V_k~\bar x_{ik} \over {\sum_{k}^{N_{\text{pixel}}}V_k}}.
\ea 
This averaging requires a choice of time slice. Since all pixels have the same surface area and hence the angular directions expand the same way today, we can assume that the radial direction also expands the same way today, which implies that the three-volume is the same today for all pixels. This amounts to identifying a volume as a pixel. 
Defining $v_{k}=V_{k}/\sum_{k}V_{k}$ and generalizing the result in \citet{rasanen_2006} for $N_{\text{pixel}}$ disjoint regions, the average of the Hubble parameter is given by
\ba
\left< {\dot a \over a}\right>_{V_k}=\sum_{k}^{N_{\text{pixel}}}v_{k} {\dot a \over a}.
\ea
Taking the derivative, it follows that 
\ba
\left<{\ddot a \over a}\right>_{V_k}=\sum_{k}^{N_{\text{pixel}}}v_{k} {\ddot a \over a} 
+2\sum_{k}^{N_{\text{pixel}}}\sum_{l>k}^{N_{\text{pixel}}}v_{k}v_{l}\left(H_{0,k}-H_{0,l}\right)^2,
\label{eqn:average_Vk}
\ea
which decomposes into a linear term in the pixel average of $(\ddot a/a)$ and a quadratic term in differences of $H_{0}$ between pairs of pixels. The quadratic (backreaction) term generates an acceleration that is due not to regions speeding up locally, but instead to the slower regions becoming less represented in the average. In the absence of the quadratic term, the volume average reduces to the pixel average above. Then the volume average of $q_{0}$ becomes
\ba
\left< q_{0}\right>_{V_k}
&=&\sum_{k}^{N_{\text{pixel}}}v_{k}q_{0,k}\cr
&-&{2\over \left(\sum_{k}^{N_{\text{pixel}}} v_{k}H_{0,k}\right)^2}
\sum_{k}^{N_{\text{pixel}}}\sum_{l>k}^{N_{\text{pixel}}}v_{k}v_{l}\left(H_{0,k}-H_{0,l}\right)^2.
\ea
Using the fluctuations in $H_{0}$ measured in the local parameter estimation, we find 
$\bar q_{0}=-0.452\pm0.159$ and $\bar q_{0,\text{flat}}=-0.586\pm0.061,$ and after the noise bias removal we find 
$\bar q_{0}=-0.527\pm0.173$ and $\bar q_{0,\text{flat}}=-0.587\pm0.083,$ (see Table \ref{table:param_pix}). 
After the noise bias removal, the volume average values become 
closer to the value for the deceleration estimated from the complete sample. The results from both averaging methods strengthen the validity of $\bar x_{ik}^{\text{bias}}$ as a measure of the noise bias due to the inhomogeneity of the SN sampling.

The quadratic term is of order $10^{-3}$ (or lower) times the linear term, which means that the error of the difference is below the standard deviation. 
Hence for the angular fluctuations in $H_{0}$ measured with this SN sample, the contribution of the quadratic term in Eq.~(\ref{eqn:average_Vk}) is insignificant, which renders the volume averaging equivalent to the pixel averaging. Hence in the context of this toy model of an inhomogeneous space-time, backreaction is not a viable dynamical mechanism to emulate cosmic acceleration.

\begin{table*}
\caption{
{\bf Values for the parameters estimated from the JLA type Ia SN sample.} 
}
\label{table:param_pix}
\centering
\begin{tabular}{c|cc|cccc}
\hline\hline
Parameter 
& \multicolumn{2}{c|} {Complete sample}
&\multicolumn{4}{c} {Subsample into pixels}
\\ 
& All & Flat 
& All & Flat & All+Unbias & Flat+Unbias 
\\ \hline
$\Omega_{M}$ 
& $0.256\pm 0.074$ & $0.267\pm 0.018$
& $0.242\pm 0.099$ & $0.276\pm 0.041$
& $0.294\pm 0.115$ & $0.256\pm0.056$ 
\\
$\Omega_{\Lambda}$ 
& $0.715\pm 0.118$ & $0.733\pm 0.018$ 
& $0.607\pm 0.115$ & $0.724\pm 0.041$
& $0.679\pm0.120$ & $0.715\pm0.056$ 
\\
$H_{0}$ 
& $71.17\pm 0.44$ & $71.21\pm 0.33$ 
& $71.06\pm 0.87$ & $70.93\pm 0.64$ 
& $71.31\pm0.94$ & $71.17\pm0.90$
\\
\hline
$\Omega_{\kappa}$ 
& $0.029\pm 0.140$ & $0.000\pm 0.026$ 
& $0.078\pm0.131$ & $0.000\pm0.058$ 
& $0.002\pm0.135$ & $0.029\pm0.079$
\\
\multirow{2}{*}{$q_0$} & \multirow{2}{*}{$-0.586\pm 0.124$} & \multirow{2}{*}{$-0.599\pm 0.020$} 
&  $-0.451\pm 0.159$ & $-0.586\pm 0.061$ 
& $-0.526\pm0.172$ & $-0.587\pm0.083$
 \\
&& 
& $-0.452\pm 0.159$ & $-0.586\pm 0.061$ 
& $-0.527\pm0.173$ & $-0.587\pm0.083$
\\
\hline
\end{tabular}
\tablefoot{
Column 1: the parameters estimated either directly or indirectly from the data. Columns 2--3: the values estimated from the complete SN sample, both when allowing $\{\Omega_{M}, \Omega_{\Lambda}, H_{0}\}$ to vary (All) and when imposing $\Omega_{M}+\Omega_{\Lambda}=1$ (Flat). Columns 4--7: the values estimated from the subsampling of SNe into pixels of equal surface area, both when allowing $\{\Omega_{M}, \Omega_{\Lambda}, H_{0}\}$ to vary (All) and when imposing $\Omega_{M}+\Omega_{\Lambda}=1$ (Flat), before and after (Unbias) the implementation of the noise bias removal. For $q_0$, the first row corresponds to the pixel averaging and the second row corresponds to the volume averaging. All the other parameters have one row corresponding to the pixel averaging.
}
\end{table*}

\section{Conclusions}
\label{sec:concl}

In this paper, we developed a method to probe the inhomogeneity of the large-scale structure and cosmic acceleration and applied to type Ia SN data. In particular, we subsampled the original SN sample into pixels of equal surface area and in each pixel we estimated the cosmological parameters that affect the luminosity distance, namely $\{\Omega_{M}, \Omega_{\Lambda}, H_{0}\},$ assuming a local Friedmann-Lema\^itre-Roberston-Walker metric in each pixel. This local parameter estimation resulted in maps of the cosmological parameters with the same pixelation as the SN subsamples. 
 
We measured fluctuations about the average values of order 
5--95\% 
for the matter energy density $\Omega_M,$ 
1--25\% 
for the dark energy density $\Omega_\Lambda$, and to 
5\% 
for the Hubble parameter $H_{0},$ which translate into fluctuations about the average values of order 
35--900\% 
for the curvature energy density $\Omega_{\kappa}$ and 1--50\% 
for the deceleration parameter $q_{0}.$  
The comparison of our results with those obtained in other SN studies indicates convergence of the results and consistency among the measurements. Owing to the spatial inhomogeneity of the SN surveys, as demonstrated by the tests performed, the pixel subsamples are statistically unequal. As a result, well-sampled pixels return robust estimations, whereas poorly sampled pixels return weak constraints. We also assessed how the local estimation can be improved by fixing one parameter. From this test, we observed that the estimation of $H_{0}$ is largely unaffected by the size of the pixel subsample, whereas the estimation of the energy densities depends highly on the size of the pixel subsample. For more robust results on $\{\Omega_{M},\Omega_{\Lambda}\},$ and subsequently on $\Omega_{\kappa},$ we would need a SN survey that covered the sky more homogeneously so that all pixels were sufficiently sampled both in number of supernovae and in redshift range.

We also computed the power spectrum of the corresponding maps of the parameters. As a result of the size of each pixel, which was selected so that all pixels contained SNe, the largest multipole accessible is the octopole $\ell=3.$ 
Previous measurements were aimed at the measurement of the dipole $\ell=1$ only. 

For an analytical toy model of an inhomogeneous ensemble of homogeneous pixels, we derived the backreaction term in the deceleration parameter due to the fluctuations of $H_{0}$ across the sky.  
The overall deceleration parameter, computed as the average over the pixels assuming a) no backreaction and b) backreaction, results in a difference of order $10^{-3}$ between the two averaging methods. This means that, for the toy model considered, the backreaction generates insignificant extra acceleration.

This paper is a proof of concept to measure deviations from homogeneity.
Despite the limitations of the SN data available to date, we demonstrated that this method can measure deviations from homogeneity in the cosmological parameters across the sky.
Investigating the averaging variability can be relevant to look for a dynamical mechanism that emulates a cosmic acceleration as a derived, instead of fundamental, effect. An idea worth exploring would be to repeat the local estimation for the dependence of other parameters without an {\it a priori} dark energy component, for example using a kinematic parametrization of the luminosity distance expressed in terms of time derivatives of the scale factor. Investigating the angular distribution of the parameters can also be relevant to characterize the pattern of overdensed regions (filaments) and underdensed regions (voids) observed on large scales. Another idea worth exploring would be to correlate the angular distribution of the cosmological parameters with probes of the large-scale structure from which a pattern of matter distribution might emerge. Yet another idea would be to use the $\sigma_8$--$H_0$ degeneracy to explore the implications of the angular variation of the cosmic expansion to neutrino physics ($\sum m_{\nu}$ or $N_{\text{eff}}$). 

Further investigation along these lines can be important for the formulation and validation of inhomogeneous models on the basis of an unprecedented amount of data that we anticipate having in the near future.\\

\noindent{\bf Acknowledgements} 
The authors thank I. Hooke, V. Marra, N. Nunes, and S. R\"as\"anen for useful discussions. CSC is funded by Funda\c{c}\~ao para a Ci\^encia e a Tecnologia (FCT), Grant no. SFRH/BPD/65993/2009.  



\appendix

\section{Estimation method}
\label{sec:est_method}

We aim to find the values $\{\Omega_{M},\Omega_{\Lambda},H_{0}\}$ that best fit $\mu_{\text{theo}}$ to $\mu_{\text{data}}.$ 
The probability that a given combination of values $\{\Omega_{M},\Omega_{\Lambda},H_{0}\}$ fits the data is known as the posterior probability $P_{\text{post}}$ and is defined as  
\ba
P_{\text{post}}(\Omega_M,\Omega_\Lambda,H_0\vert \{\mu_{\text{data}}(z_i)\})
={P(\{\mu_{\text{data}}(z_i)\}\vert \Omega_M,\Omega_\Lambda,H_0) \over P(\{\mu_{\text{data}}(z_i)\}}
P_{\text{prior}}.
\!\!\!\!\!\!\!\!\!\cr
\ea
The probability of the data given the values for the model parameters (the numerator) is known as the likelihood, and the integral of the likelihood over the parameters' space (the denominator) is known as the evidence. The prior probability $P_{\text{prior}}$ describes informed constraints  that we impose {\it a priori} on the parameters.

We assume that the measurements of $\mu$ at the different $z$s are independent and that the noise is Gaussian as a first approximation. Then the likelihood is given by
\ba
&&{\cal L}(\Omega_M,\Omega_\Lambda,H_0) 
\equiv 
P(\{\mu_{\text{data}}(z_i)\}\vert \Omega_M,\Omega_\Lambda,H_0)\cr
&=&{1\over {(2\pi)^{N_{\text{SNe}}/2} } \prod^{N_{\text{SNe}}}_{i}\sigma_{\mu_{\text{data}}}(z_i)}\\
&&\times\exp\left[ -{1\over 2}\sum^{N_{\text{SNe}}}_{i}
\left( {\mu_{\text{data}}(z_i) -\mu_{\text{theo}}(z_i;\Omega_{M},\Omega_{\Lambda}, H_{0}) } 
\over {\sigma_{\mu_{\text{data}}}(z_i)}\right)^2
\right].
\ea
Each combination of values $\{\Omega_{M},\Omega_{\Lambda}, H_{0}\}_{\text{trial}}$ is generated from a Gaussian distribution and defines a trial. We  implement the Metropolis-Hastings algorithm to produce Markov chains where the samples are generated from the conditional posterior distribution with the following priors:
a)  $\Theta(\Omega_{M})\Theta(\Omega_{\Lambda})\Theta(H_{0})$ and 
b) $\Theta(1-\Omega_{M})\Theta(1-\Omega_{\Lambda}),$ with $\Theta$ standing for the Heaviside function.
For each trial we define
\ba
\chi^2_{\text{trial}}=\sum^{N_{\text{SNe}}}_{i}
\left( {\mu_{\text{data}}(z_i) -\mu_{\text{theo}}(z_i;\{\Omega_{M},\Omega_{\Lambda}, H_{0}\}_{\text{trial}})}
\over {\sigma_{\mu_{\text{data}}}(z_i)}\right)^2,
\ea
and compare the likelihood between the trial combination and the previous combination by taking the ratio of the corresponding likelihoods times the prior as follows:
\ba
LR=\exp\left[-{1\over 2}(\chi^2_{\text{trial}}-\chi^2_{\text{prev}})\right] 
{P_{\text{prior}}(\{\Omega_{M},\Omega_{\Lambda}, H_{0}\}_{\text{trial}}) \over 
P_{\text{prior}}(\{\Omega_{M},\Omega_{\Lambda}, H_{0}\}_{\text{prev}}) }.\ea
If the ratio is larger than one, then the trial combination is accepted and added to the chain. Otherwise, the ratio is compared with a trial from a uniform distribution defined in $[0,1],$ and if the ratio is larger, then the trial combination is accepted and added to the chain; otherwise, the trial combination is rejected and the previous combination is added to the chain instead.



\end{document}